\documentclass[prb,superscriptaddress]{revtex4}
\usepackage{amsmath}
\usepackage{graphicx}
\usepackage{amssymb}

\makeatletter

\providecommand{\tabularnewline}{\\}

\begin{document}
\author{Alexey A. Polotsky}
\affiliation{Institute of Macromolecular Compounds of the Russian Academy of Sciences,
31 Bolshoy pr., 199004 Saint-Petersburg, Russia}
\email{alexey.polotsky@gmail.com}
\author{Elizaveta E. Smolyakova}
\affiliation{Saint-Petersburg State University, Physical Faculty, 1 Ulyanovskaya
ul., 198504 Petrodvorets, Saint-Petersburg, Russia}
\author{Tatiana M. Birshtein}
\affiliation{Institute of Macromolecular Compounds of the Russian Academy of Sciences,
31 Bolshoy pr., 199004 Saint-Petersburg, Russia}

\title[Theory of mechanical unfolding of homopolymer globule]
{Theory of mechanical unfolding of homopolymer globule: all-or-none transition in force-clamp mode vs phase coexistence in position-clamp mode}

\begin{abstract}
Equilibrium mechanical unfolding of a globule formed by long flexible
homopolymer chain collapsed in a poor solvent and subjected to an
extensional force $f$ (force-clamp mode) or extensional deformation
$D$ (position-clamp mode) is studied theoretically. Our analysis,
like all previous analysis of this problem, shows that the globule
behaves essentially differently in two modes of extension. In the
force-clamp mode, mechanical unfolding of the globule with increasing
applied force occurs without intramolecular microphase segregation,
and at certain threshold value of the pulling force the globule unfolds
as a whole ({}``all-or-none'' transition). The value of the threshold
force and the corresponding jump in the distance between the chain
ends increase with a deterioration of the solvent quality and / or
with an increase in the degree of polymerization. In the position-clamp
mode, the globule unfolding occurs via intramolecular microphase coexistence
of globular and extended microphases followed by an abrupt unraveling
transition. Reaction force in the microphase segregation regime demonstrates
an {}``anomalous'' decrease with increasing extension. Comparison
of deformation curves in force and position-clamp modes demonstrates
that at weak and strong extensions the curves for two modes coincide,
differences are observed in the intermediate extension range. Another
unfolding scenario is typical for short globules: in both modes of
extension they unfold continuously, without jumps or intramolecular
microphase coexistence, by passing a sequence of uniformly elongated
configurations. The values of the the critical chain length, $N_{cr},$
separating long and short chain behavior are slightly different for
two extension modes: $N_{cr,f}<N_{cr,D}.$
\end{abstract}
\maketitle

\section{Introduction}

In the last decade, a substantial progress is achieved in development
of experimental single molecular manipulation techniques. The use
of such instruments as atomic force microscope (AFM) \cite{Forman:2007,Borgia:2008,Puchner:2009},
optical tweezers \cite{Williams:2002,Moffitt:2008} and magnetic traps
\cite{Meglio:2009} made it possible to study the mechanical properties
and related conformational changes in individual (bio)macromolecules
and their complexes or the dynamics and mechanisms of molecular motors.
The above mentioned methods not only became {}``standard'' tools
of biophysics, polymer physics, and molecular biology but also led
to appearance of a new interdisciplinary field of molecular nanomechanics.

The choice of the proper experimental approach depends on the system
and the problem under study. Each apparatus has its effective range
of applied (and measured) forces and extensions and, as a rule, more
than one operating modes. For example, in AFM experiment the ends
of the macromolecule under study are bound to AFM cantilever tip and
to a flat surface that are moving apart. The basic operating mode
of AFM is the \emph{velocity-clamp mode} when the surface and the
AFM tip holder are moved relative to one other at constant velocity.
On the other hand, using a feedback loop it is possible to control
the force applied to investigated object and carry on measurements
either in the \emph{force-clamp mode}, where applied force is hold
constant or in the \emph{force ramp mode} where the force increases
linearly with time \cite{Samori:2005,Oberhauser:2001,Schlierf:2007}.
AFM has a wide effective range of measured and applied forces: 10
pN to 100 nN, The basic operating mode of optical tweezers is the
\emph{position-clamp mode} where extension is fixed with the aid of
one or two optical traps. However the use of a feedback loop allows
to operate in the force-clamp mode too. The effective force range
of optical tweezers is typically 0.1 - 150 pN. A magnetic trap consists
of a set of magnets that provide a strong magnetic field gradient,
thus exerting a force on magnetic beads tethered by a macromolecule
of interest. For this set-up it is most natural and convenient to
study macromolecules subjected to a constant applied force (force-clamp
mode). At the same time magnetic traps has a high resolution that
allows to measure the response of macromolecule to relatively weak
forces about or less than 1 pN. On the other hand, the maximum force
can be increased (up to 160 pN) by using bigger beads \cite{Meglio:2009}.
It should be also mentioned that magnetic traps and optical tweezers
allow studying rotational degrees of freedom of single macromolecules
by applying (or measuring) a torque on them \cite{Moffitt:2008,Meglio:2009}.
This feature is especially useful and informative in studying DNA
and DNA-protein complexes .

Hence, different modes of mechanical action on single macromolecules
are available in experiment. These modes fall far short of being equivalent
with respect to the results obtained. Results of single-molecular
nanomechanical experiments in the position $(D)$ and force $(f)$
clamp modes are expressed in the form of equilibrium deformation curves
$f=f(D)$ in the former case and $D=D(f)$ in the latter case. Equilibrium
deformation curves $f=f(D)$ and $D=D(f)$ can also be obtained from
the velocity-clamp and force ramp experiments, respectively, if the
corresponding governing parameter is changed very slowly. 

From the statistical-mechanical point of view, position-clamp and
force-clamp modes can be interpreted as two conjugate thermodynamic
\emph{ensembles}: fixed extension ensemble, or $D$-ensemble, and
fixed force ensemble, or $f$-ensemble, respectively. Correspondingly,
in the following text both the words {}``mode'' and {}``ensemble''
will be utilized as {}``equal in rights''. According to classical
statistical mechanics \cite{Balescu:1975} conjugate $D$- and $f$-
ensembles are equivalent, i.e. corresponding deformation curves $f=f(D)$
and $D=D(f)$ should coincide, it is enough to exchange axes on one
of the dependences, for instance $D(f)\to f(D)$. However, statistical
mechanics deals with macroscopic systems. For individual macromolecules
this limit (which is called {}``thermodynamic limit'') is achieved
when the chain length $N$ tends to infinity $N\to\infty$. Macromolecules
with finite chain length represent small systems (or nanosystems),
therefore, their behavior in different thermodynamic ensembles should
be individually studied \cite{Skvortsov:2009}. 

\begin{figure}[t]
\includegraphics[width=10cm]{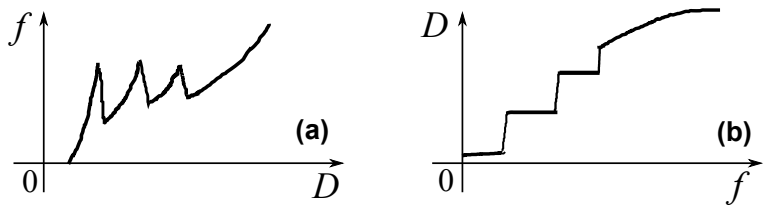}

\caption{\label{fig:schem_dc_protein}Schematic pictures of deformation curves
for globular protein mechanically unfolded in (a) position-clamp and
(b) force-clamp modes.}

\end{figure}

Experimental force-extension dependences $f=f(D)$ obtained for stretching
of macromolecules in the position-clamp mode (or in the velocity clamp
mode at very low velocity) have a rather complex structure. For example,
for DNA double helix, the reaction force $f$ grows with an increase
in the extension $D$ at small and at large extensions. Two increasing
branches are separated by a plateau at moderate extensions. Constant
force on the plateau is an indication of progressive extension-induced
melting of the native double helix structure of DNA \cite{Williams:2002}.
Mechanical action on different parts of protein globules (or computer
simulation of such an action) leads to appearance of a single or multiple
maxima on the force-extension curve followed by decrease(s) in reaction
force with increasing extension \cite{Forman:2007,Borgia:2008,Puchner:2009},
Fig.~\ref{fig:schem_dc_protein} a. Each peak of this {}``sawtooth''
curve corresponds to the unfolding of an individual domain of the
protein.

In the force-clamp mode, deformation curves of globular proteins acquire
appreciably different {}``staircase'' form demonstrating a sequence
of jump-wise extensions when applied force is increased \cite{Oberhauser:2001,Schlierf:2007},
 Fig.~\ref{fig:schem_dc_protein} b. Each step of this {}``staircase''
can also be associated with individual domains' unfolding.

A globule formed by a homopolymer chain collapsed in a poor solvent
is similar to a liquid droplet and has, therefore, a simpler structure
that that of the protein globule which is, according to Schr\"odinger
\cite{Schroedinger:1944}, an {}``aperiodic crystal''. Nevertheless,
as it follows from the theory \cite{Halperin:1991:EL,Halperin:1991:MM,Cooke:2003,Craig:2005:1,Polotsky:2009,Polotsky:2010:MM},
the shape of force-extension curves in the position-clamp mode for
homopolymer globules were found to be not so simple as it might be
expected. 

\begin{figure}[t]
\includegraphics[width=7cm]{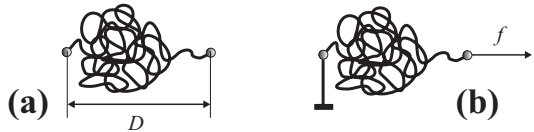}

\caption{\label{fig:Globule}Homopolymer globule deformed in the position-clamp
(a) and force-clamp (b) mode}

\end{figure}

The goal of the present work is to develop a theory of mechanical
unfolding in the force-clamp mode of a globule formed by a homopolymer
chain collapsed in a poor solvent, Fig.~\ref{fig:Globule}. Comparison
with available data on homopolymer globule unfolding in the position-clamp
mode should allow revealing main regularities that determine differences
in globule behavior in the the $f$- and the $D$-ensembles. 

The problem of the homopolymer globule unfolding attracts the attention
of theorists starting from the pioneering work of Halperin and Zhulina
\cite{Halperin:1991:EL} in 1991. The following studies included the
development of analytical theories and computer simulations, a review
of these works can be found, for example, in \cite{Skvortsov:2009}
and \cite{Polotsky:2009,Polotsky:2010:MM}. Recently we have performed an extensive
self-consistent field (SCF) modeling of the globule unfolding in the
position clamp-mode \cite{Polotsky:2009} and proposed a quantitative
mean-field theory based on simple model \cite{Polotsky:2010:MM} wich
shown a very good agreement with the results obtained by SCF modeling.
By now a clear theoretical
picture of equilibrium unfolding of the polymer globule by extensional
deformation is well developed.

\begin{figure}[t]
\includegraphics[width=12cm]{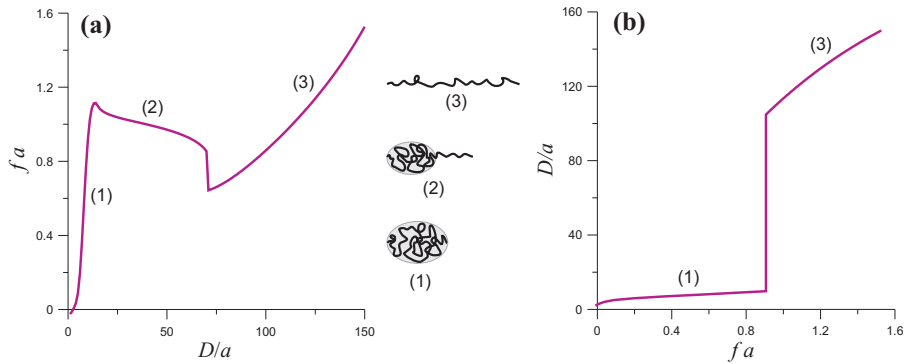}

\caption{\label{fig:FEC_and_conf}Homopolymer globule deformation curves calculated
at $N=200$, $\chi=1.4$ from SCF modeling in the position-clamp (a)
and force-clamp (b) modes.}

\end{figure}

Consider a flexible polymer chain comprising $N$ monomer units, each
of size $a$, immersed in a poor solvent. The solvent strength is
characterized by the Flory-Huggins parameter $\chi>0.5$. A typical
equilibrium force-extension curve for (not too small) globule unfolded
by extensional deformation (in the position-clamp mode) is presented
in  Fig.~\ref{fig:FEC_and_conf} (a). Here and below $k_{B}T$ is taken
as energetic unit (hence, $fa$ is dimensionless). At small deformations
the globule extends as a whole and has an elongated shape - this corresponds
to the first ascending part of the curve (1). Further increase in
the end-to-end distance leads to appearance of a new intramolecular
microphase, i.e. a (strongly) stretched chain part. Microphase segregation
in the extended globule has an analogy with the Rayleigh instability
in a liquid droplet \cite{Rayleigh:1882}; it was predicted in the
first work on globule unfoldung by Halperin and Zhulina \cite{Halperin:1991:EL}.
In a wide range of extensions, a microphase segregation takes place,
and the globule acquires a “tadpole” conformation with a prolate globular
“head” and a stretched “tail”. On the force-extension curve this regime
corresponds to the weakly decreasing quasi-plateau (2). The deformation
of the tadpole is accompanied by progressive unfolding of the globular
core, at strong deformations the tadpole conformation becomes unstable,
and at certain extension the globule completely unfolds to a uniformly
extended chain. This transition is accompanied by a sharp decrease
in the reaction force, the number of monomer units in the vanishing
globular phase is rather large: $n\sim N^{3/4}$. This ``unraveling''
transition was discovered by Cooke and Williams \cite{Cooke:2003}
in 2003. The second ascending part of the force-extension curve (3)
characterizes the following extension of the unfolded chain.

Three factors have an impact on such a complex picture of globule
unfolding. First, unfolding of the globule occurs as an intramolecular
phase transition. Second, the globule is a small nano-sized system
(nanosystem). In contrast to (infinitely) large macroscopic systems,
the fraction of elements (i.e. monomer units in a macromolecule) at
the interface is not negligible and grows with the chain extension.
Finally, the macromolecule has a {}``linear memory'' (I.M. Lifshitz
\cite{Lifshitz:1968}): all elements are joined into a single chain
at any extension $D$ that cannot exceed the contour length $Na$.

The present paper aims at the developing a mean-field\emph{ }theory
of equilibrium globule unfolding in the force-clamp mode (which means
that the kinetic aspects of globule lie out of scope of the present
study). We will show that the dependence $D=D(f)$ differs fundamentally
from $f=f(D)$. As it can be seen from  Fig.~\ref{fig:FEC_and_conf} b,
$D=D(f)$ curve is free from peculiarities of the $f=f(D)$ curve
in  Fig.~\ref{fig:FEC_and_conf} a and has a simple shape characterized
by a sharp transition from the globular to the completely unfolded
{}``open chain'' state. 

Note that equilibrium deformation curves shown in  Fig.~\ref{fig:FEC_and_conf}
represent the dependences of the \emph{average} reaction force (end-to-end
distance) on imposed deformation (applied force), hence, strictly
speaking, one should write: $\left\langle f\right\rangle =\left\langle f(D)\right\rangle $
($\left\langle D\right\rangle =\left\langle D(f)\right\rangle $).
In the following text, angular brackets at observables will be omitted.
This should not confuse the reader because the name of the argument
(as well as the name of the function) in the dependence definitely
points on the corresponding mode of deformation (thermodynamic ensemble).

It is important to emphasize that in the present work, the problem
is solved in the framework of the \emph{mean-field} approach. It is
well-known that the mean-field approach neglects fluctuations around
the ground state for the considered system, see for example \cite{GrosbergKhokhlov:1994}.
This approximation is justified for large systems, except for the
vicinity of the phase transition point where mean field theory predicts
a jumpwise transition. Neglecting fluctuations in the case of the
globule unfolding does not affect, however, the essential physics:
as it will be demonstarated below, the main ``source'' of the differences
in globule unfolding in position- and force-clamp modes is the surface
energy of the globule.

The rest of the paper is organized as follows. In Section {}``Self-consistent
field modeling of globule extension in position- and force-clamp modes''
 we show how to use the results of SCF modeling obtained for the position-clamp
mode ($D$-ensemble) in order to derive deformation curves in the
force-clamp mode ($f$-ensemble) without doing additional extensive
calculation and compare the deformation curves in two modes. In Section
{}``Model and analytical theory of globule unfolding in the $f$-ensemble''
analytical theory of the globule unfolding in the $f$-ensemble based
on a simple model is developed and the results of theory, including
detailed study of the force-induced unfolding transition and comparison
between unfolding transition in two ensembles are presented, and this
is followed by Discussion and Conclusions.

\section{\label{sec:SCF}Self-consistent field modeling of globule extension
in position- and force-clamp modes}

\subsection{Relation between $D$- and $f$-ensembles and {}``translation''
of results}

As it was already mentioned in the Introduction, position-clamp and
force-clamp modes of globule extension correspond to two conjugate
thermodynamic $D$- and $f$- ensembles, respectively. It is well
known that each thermodynamic ensemble has its {}``own'' independent
variables and this determines a choice of the proper thermodynamic
potential that describes the thermodynamics of the system in the given
ensemble. For the fixed extension ($D$-) ensemble the proper thermodynamic
potential is the Helmholtz free energy $F(D)$, while for the fixed
force ($f$-) ensemble one should work with the Gibbs free energy
$G(f)$. These thermodynamic potentials are related via Laplace transform
\cite{Landau:5,Sinha:2005,Winkler:2010} :

\begin{equation}
G(f)= - k_{B}T\:\log\left[\int dD\, e^{-(F(D)-fD)/k_{B}T}\right]\label{eq:GF_Laplace}\end{equation}
where the integral is taken over all possible extensions (end-to-end
distances) $D$, therefore, fluctuations of the chain end positions
are taken into account. In the self-consistent field
approach one neglects fluctuations and considers only one state that
maximizes the exponent in Eq. (\ref{eq:GF_Laplace}), i.e. the integral
is dominated by the saddle point value \cite{GrosbergKhokhlov:1994}
and, therefore, the thermodynamic potentials are related by Legendre
transform:

\begin{equation}
G=F-f\cdot D\label{eq:GF_Legendre}\end{equation}

\begin{figure}[t]
\includegraphics[width=14cm]{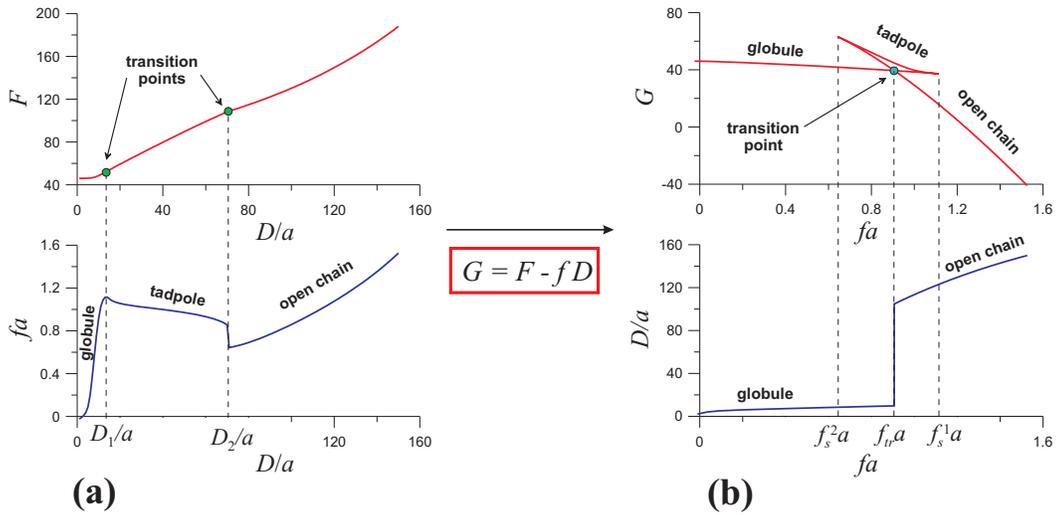}

\caption{\label{fig:Legendre} {}``Translation'' of the results obtained
for globule unfolding in the $D$-ensemble using SCF modeling into
the $f$-ensemble: using the Legendre transform (see explanation in
the text). $N=200$, $\chi=1.4$.}

\end{figure}

This relation (\ref{eq:GF_Legendre}) suggests an easy and elegant
way to transform the $D$-dependences of the Helmholtz free energy
and the reaction force in the $D$-ensemble obtained earlier in \cite{Polotsky:2009}
using Scheutjens-Fleer self-consistent field (SF-SCF) numerical approach,
into $f$-dependences of the Gibbs free energy and extension in the
$f$-ensemble. In other words, without doing independent extensive
SCF calculations for the globule unfolding in the force-clamp mode,
the results for this ensemble can be obtained by using the already
existing results for the conjugate ensemble and the Legendre transform,
Eq. (\ref{eq:GF_Legendre}). The scheme and the result of this {}``translation''
for $N=200$ and $\chi=1.4$ are presented in   Fig.~\ref{fig:Legendre}
(only the equilibrium $F(D)$ and $f(D)$ dependences are used for
such the {}``translation'').

The dependence of the Gibbs free energy obtained by {}``translation''
of the results from $D$- to $f$-ensemble has a loop,   Fig.~\ref{fig:Legendre}
b. Each part of this loop-like dependence is unambiguously identified
with the corresponding conformation of the globule (ellipsoidal globule,
tadpole, stretched chain), by analogy with the force-extension curve
for the $D$-ensemble. We see that the lowest free energy always corresponds
to one of two pure states: weakly extended globule at small forces
and stretched {}``open'' chain at large applied forces. Two free
energy branches corresponding to pure states intersect at certain
value of $f=f_{tr}$ which is the unfolding transition point. The
transition point does not coincide with none of the two characteristic
points in the $D$-ensemble, forces $f_{s}^{1}$ and $f_{s}^{2}$
corresponding to extensions $D_{1}$and $D{}_{2}$ in the $D$-ensemble
but lies between them (note also that $D_{1}$and $D{}_{2}$ are the
points bounding the range of the tadpole (two-phase) state stability,
   Fig.~\ref{fig:Legendre} a).    Fig.~\ref{fig:Legendre} b demonstrates that
the free energy of the microphase segregated tadpole state is larger
than that of pure states at any value of applied force $f$. This
points to disadvantage of the microphase segregation in the $f$-ensemble.
Extension-force curves in the force-clamp mode,    Fig.~\ref{fig:Legendre}
b reproduce the {}``all-or-none'' phase transition from weakly extended
globule to the open chain state.

\subsection{\label{sub:SCF:FE_and_DC}Free energy and deformation curves}

The above described $D-$to $-f-$ensemble {}``translation'' was
made for the SCF data obtained for different chain lengths $N=100,\:200,\:500$
and a series of Flory interaction parameters $\chi=0.8,\:1.0,\:\ldots\:,\:2.0$.
The SCF results for globule unfolding in the $D$-ensemble $N=200$
and 500 were presented in \cite{Polotsky:2009}, the results for $N=100$
was calculated in the present work.

\begin{figure}[t]
\includegraphics[width=8cm]{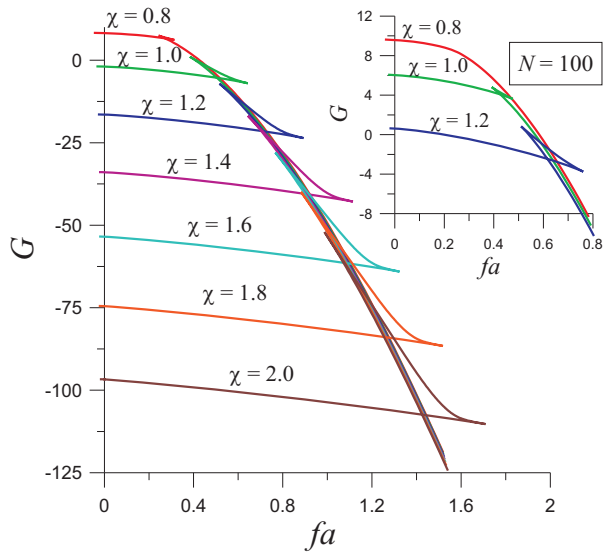}

\caption{\label{fig:G_vs_force_SCF_200}Gibbs free energy of the globule extended
by applied force obtained from SCF modeling at $N=200$ and 100 (inset)
and different $\chi$ values.}

\end{figure}

In    Fig.~\ref{fig:G_vs_force_SCF_200} the Gibbs free energy as a function
of applied force is shown for $N=100$ and 200 at different values
of $\chi$ . Most of the dependences have the loop-like shape, the
branches corresponding to open chain state collapse into a single
curve. Extended globule branches demonstrate close to linear dependence.
With an increase in $\chi$ the globule branch shifts down so that
the point of its intersection with the open chain branch (i.e. the
transition point) moves toward larger values of applied force $f$.
Unstable parts of loops at $N=200$, $\chi\geq1$ and $N=100$, $\chi\geq1.2$
in   Fig.~\ref{fig:G_vs_force_SCF_200} correspond to the two-phase tadpole
conformation which is the stable state of the globule unfolded in
the $D-$ensemble. On the other hand, at $N=200$, $\chi=0.8$ and
$N=100$, $\chi=1.0$ globule unfolding in the $D-$ensemble is not
accompanied by microphase segregation and formation of the tadpole
state \cite{Polotsky:2009} since the system is below the critical
point, i.e. the chain length $N$ is lower than the critical value
$N_{cr}(\chi)$, the minimal chain length at which the formation of
the tadpole structure is possible \cite{Polotsky:2010:MM}. Under
pre-critical conditions ($N<N_{cr}$), intermediate state in the $D-$ensemble
is that of the {}``sparse'' globule: upon stretching the globule
becomes more asymmetric and the density of its core decreases. However,
the {}``anomalous'' decrease of the reaction force $f$ with an
increase in $D$ retains in this regime. Unstable part of the loop
in   Fig.~\ref{fig:G_vs_force_SCF_200} at pre-critical values of parameters
correspond to {}``sparse'' globule conformation. At $N=100$, $\chi=0.8$
when even the unperturbed globule is on the edge of stability, the
transition becomes continuous. 

\begin{figure}[t]
\includegraphics[width=7cm]{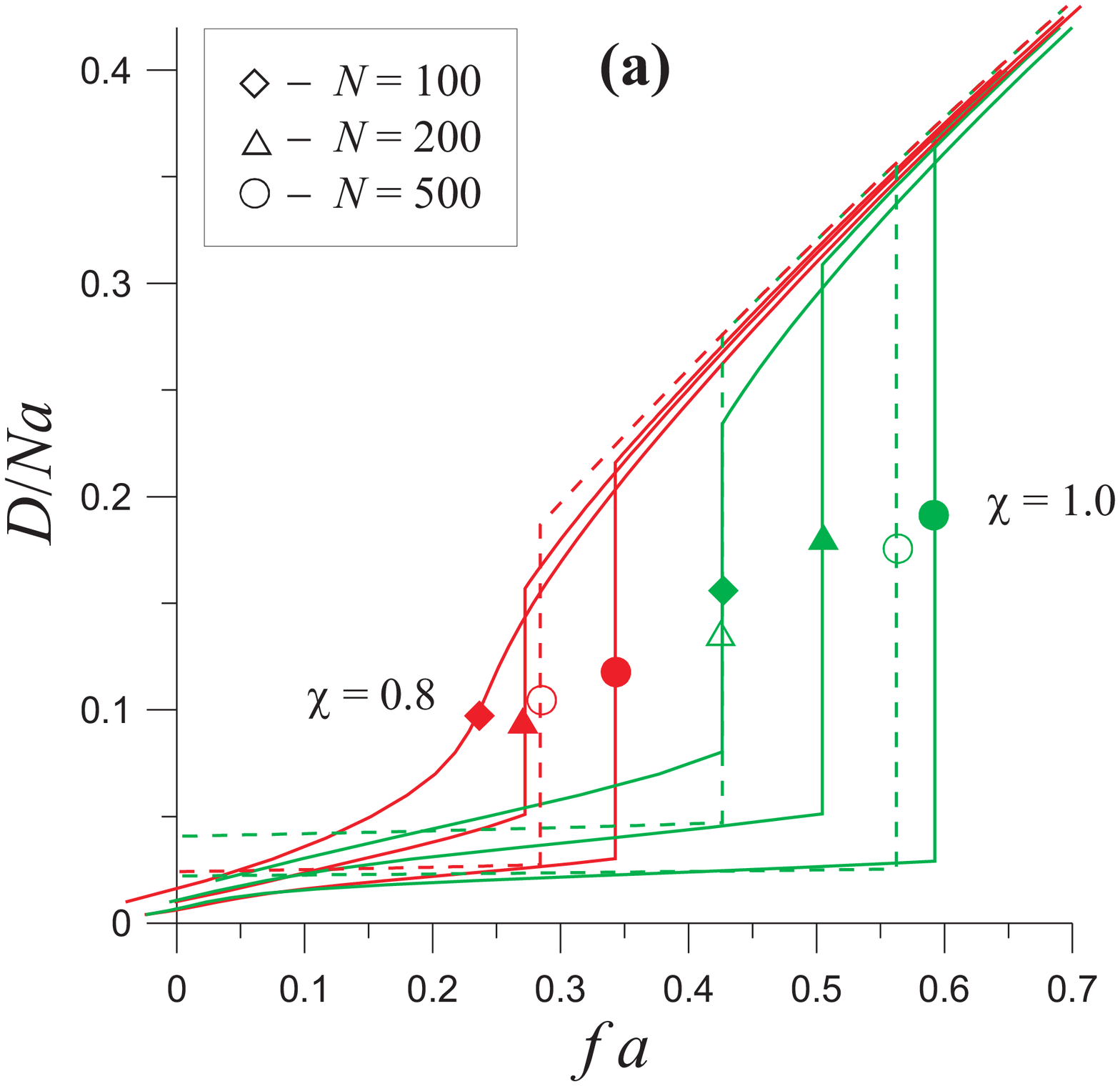} \includegraphics[width=7cm]{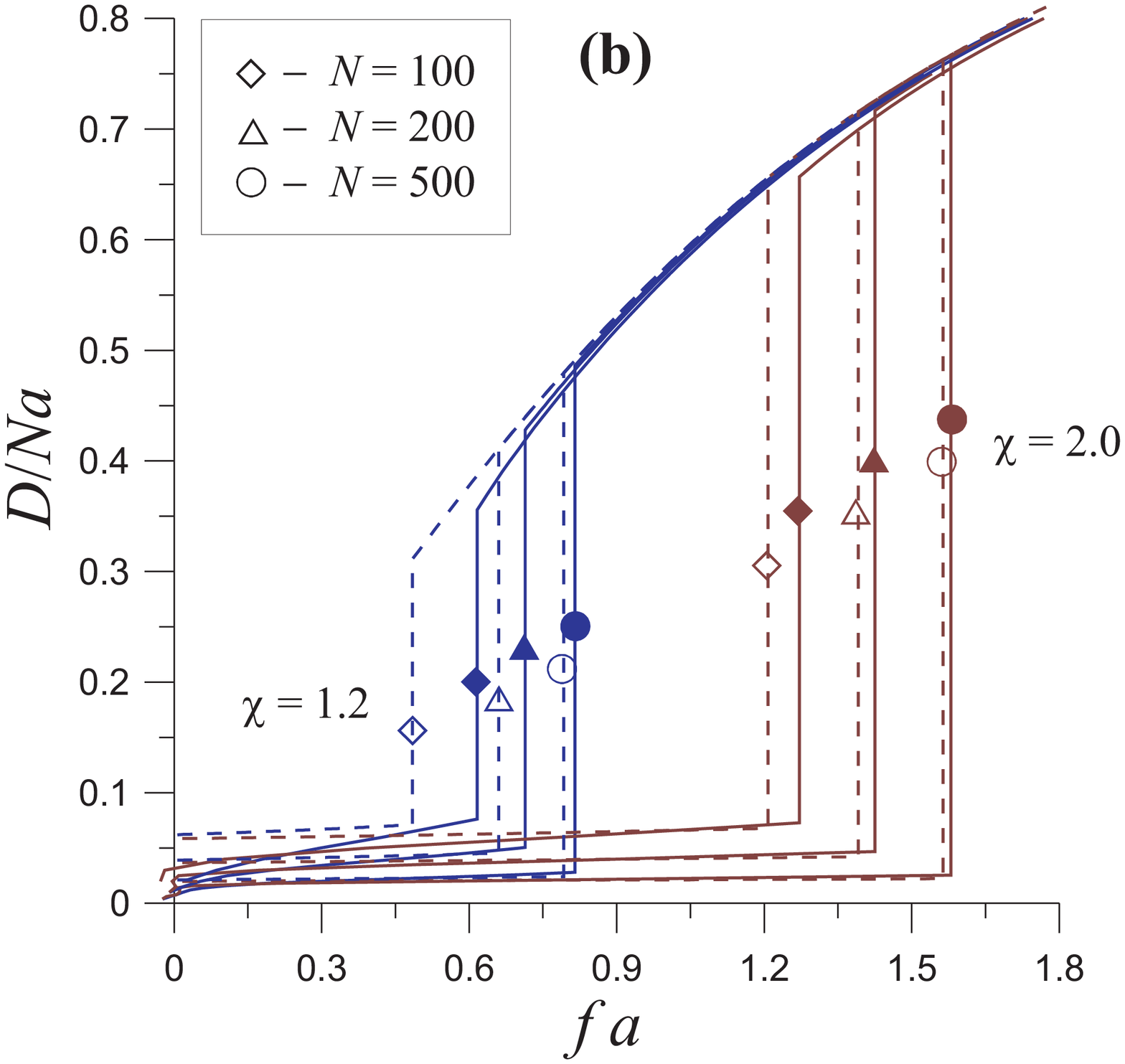}

\caption{\label{fig:D_vs_f_reduced} Equilibrium degree of extension (relative
extension) vs applied force curves for the globules with different
$N$ and $\chi$ calculated from SCF modeling data (solid lines) and
using analytical theory (dashed lines). }

\end{figure}

Deformation (extension-force) curves obtained for different values
of $N$ and $\chi$ from SCF modeling data according to the above
described procedure (see Fig.~\ref{fig:Legendre}) are presented in Fig.~\ref{fig:D_vs_f_reduced}
by solid curves. Note that for the sake of comparison, it is more
convenient to consider the relative extension, or degree of stretching,
$D/(Na)$ rather than its absolute value $D$. It can be seen that
in all presented examples, with the exception of the case $N=100$,
$\chi=0.8$, we obtain that globule unfolding in the $f$-ensemble
occurs according to the {}``all-or-none'' principle, or as a jumpwise
first order phase transition. At small applied force there is a weakly
extended globular phase, its relative linear size $D/(Na)$, weakly
increasing with growing $f$, decreases with an increase in $N$ and
$\chi$. At certain value of $f=f_{tr}$ corresponding to the transition
point the globule disintegrates and the macromolecule acquires a strongly
stretched {}``open'' conformation with the end-to-end distance $D$
proportional to $N$ but independent of $\chi$: relative extension
vs. force dependence above the transition point ($f>f_{tr}$) is described
by a unique curve independent of both $N$ and $\chi$. The position
of the transition point $f_{tr}$ where the end-to-end distance abruptly
changes depends primarily on $\chi$ and in a less extent on $N$:
$f_{tr}$ increases with increasing $N$ and $\chi$; correspondingly,
the value of the jump in the relative end-to-end distance in the transition
point increases too.

As opposed to the above considered jumpwise transition, for the smallest
$N$ and $\chi$ values presented in Fig.~\ref{fig:D_vs_f_reduced}, $N=100$,
$\chi=0.8$, globule unfolding in the $f-$ensemble occurs continuously.

\subsection{Comparison of deformation curves in force- and position-clamp modes}

One of the most important goals of the present study is to compare
the deformation curves in force- and position-clamp modes of globule
deformation. Making such a comparison is easy: the deformation curves
in two modes should be plotted together in the $D-f$ coordinates
(therefore, we exchange the axes for deformation curves in the $f-$ensemble).

\begin{figure}[t]
\includegraphics[width=7cm]{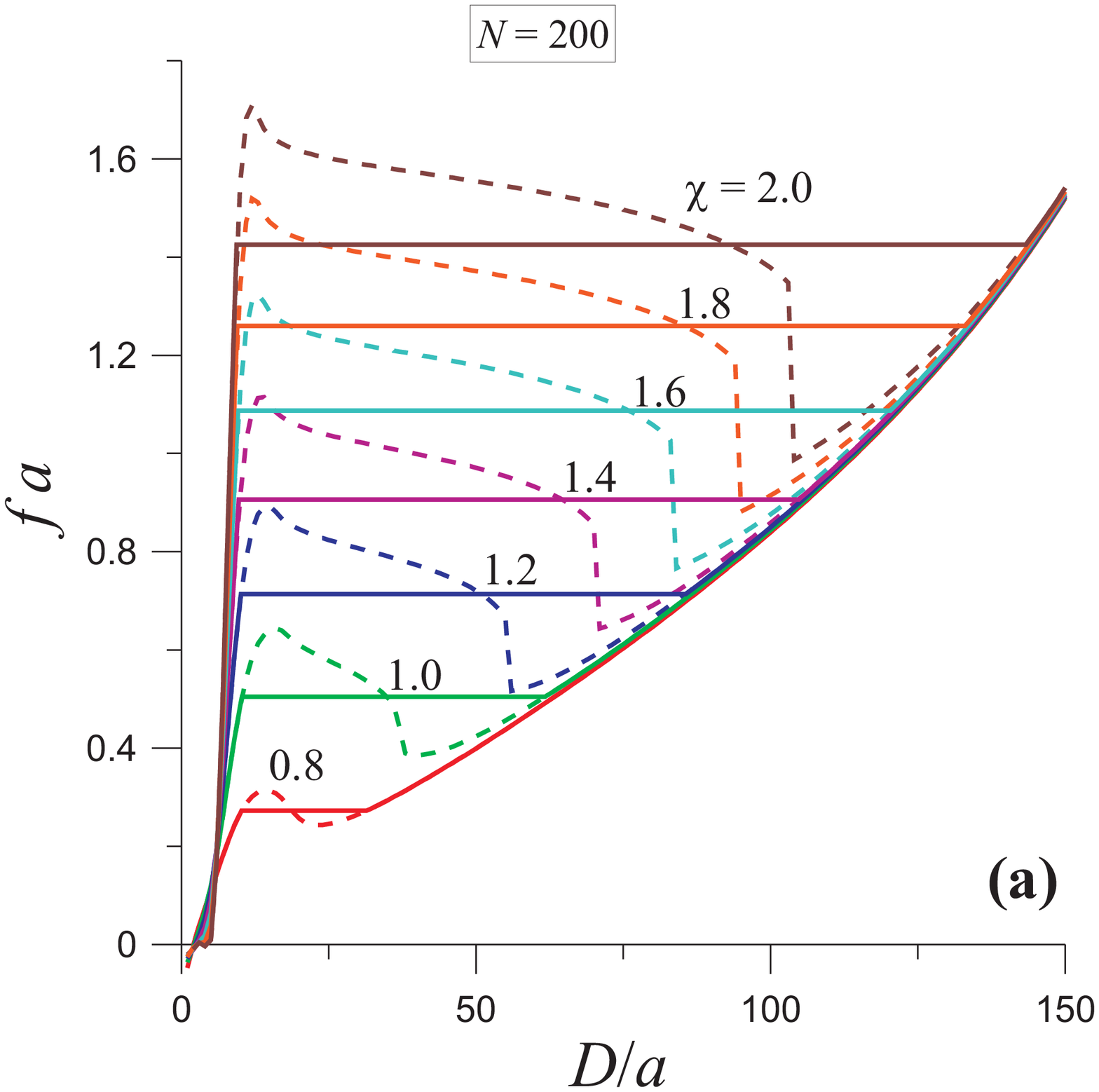} 
\includegraphics[width=7cm]{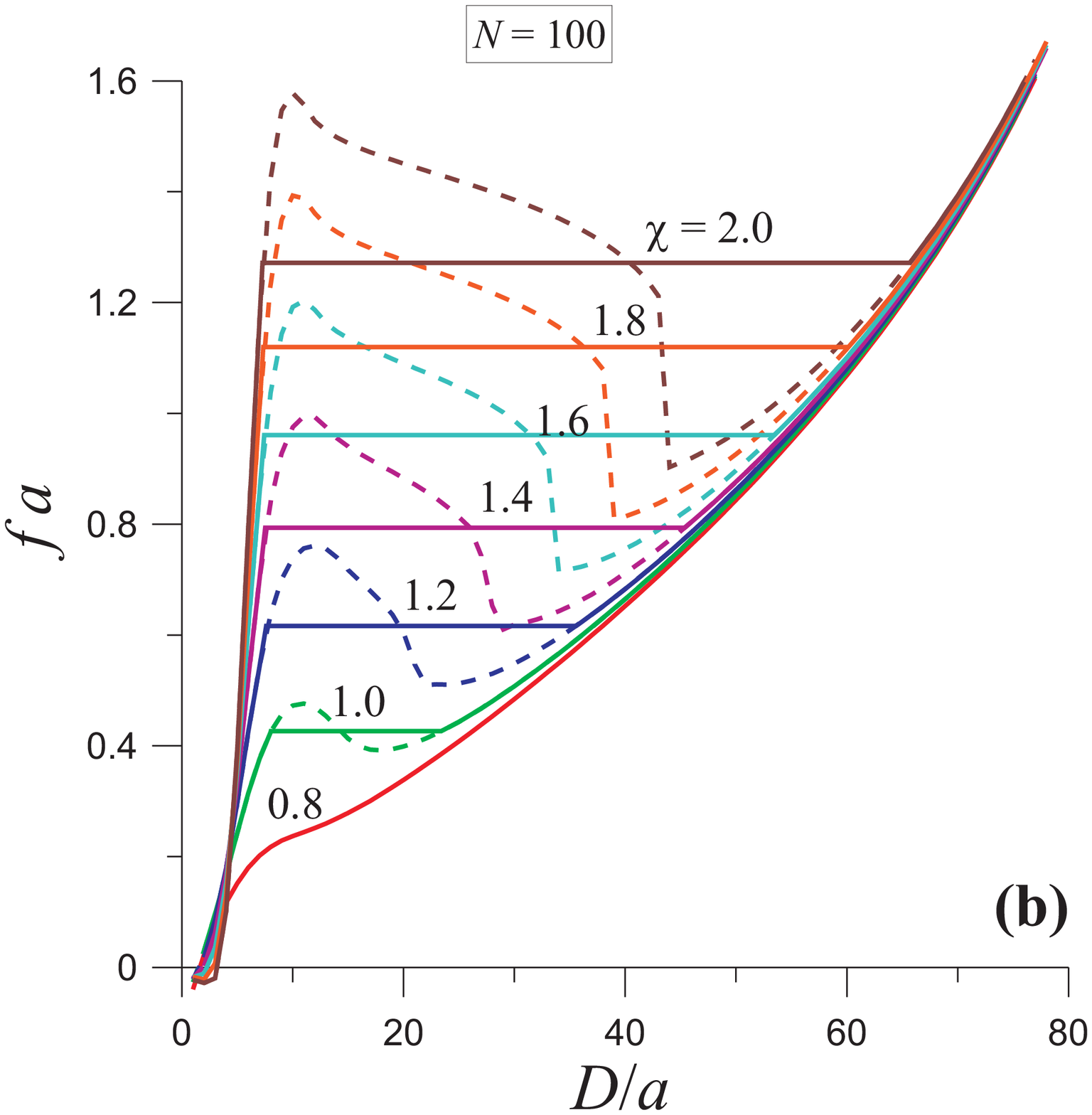}

\caption{\label{fig:comp_f_vs_D}Comparison of deformation curves for $f$-
(solid lines) and $D$-ensembles (dashed line) calculated using SCF
modeling data for $N=200$ (a), $N=100$ (b) and different values
of $\chi$.}

\end{figure}

Fig.~\ref{fig:comp_f_vs_D} presents equilibrium deformation curves for
$D-$ and $f-$ensembles calculated using SF-SCF approach for $N=200$
(Fig.~\ref{fig:comp_f_vs_D} a) and 100 (Fig.~\ref{fig:comp_f_vs_D} b) ($f-$ensemble
- solid lines, $D-$ensemble - dashed lines). Deformation curves for
the $D-$ensemble were already presented in \cite{Polotsky:2009}.
At weak and strong extensions the curves for two ensembles coincide,
differences are observed in the intermediate extension range. In this
range of extensions, at large enough values of $N$ and $\chi$, microphase
segregated tadpole structure forms and the force-extension dependence
has the quasi-plateau shape with {}``anomalous negative'' slope
and a jump in the point where the globular phase disappears. In the
cases $N=200$, $\chi=0.8$ (Fig.~\ref{fig:comp_f_vs_D} a) and $N=100$,
$\chi=1\text{.}0$ (Fig.~\ref{fig:comp_f_vs_D} b) - short macromolecule
in a moderately poor solvent - intramolecular microphase segregation
does not occur but the globule deforms as a whole and its density
decreases (the system is below the critical point \cite{Polotsky:2009,Polotsky:2010:MM}),
at the same time the deformation curve in the $D-$ensemble has a
part with the {}``anomalous'' negative slope. 

When the globule is deformed by applied force, equilibrium intermediate
state (either the tadpole or sparse globule) does not form, i.e. the
globule {}``hurdles'' these intermediate states. Horizontal plateaus
on the dependences in   Fig.~\ref{fig:comp_f_vs_D} are virtual since they
correspond to the jump in the end-to-end distance only.

Note that the quasi plateaus with {}``anomalous'' decay of the reaction
force with increasing $D$ in the $D-$ensemble may seem to be a sign
of instability. However, in the position-clamp mode they do represent
stable equilibrium states. Their {}``anomalous'' shape is caused
by the small size of the (nano)system. Nevertheless, it can be shown
that the value of threshold force $f_{tr}$ that induces the phase
transition in the force-clamp mode can be found from the deformation
curves obtained in the position-clamp mode with the aid of the Maxwell
equal area rule \cite{Landau:5} which is commonly used, to exclude
unstable states. Indeed, in the transition point $G_{globule}(f_{tr})=G_{chain}(f_{tr})$.
On the other hand, using the Legendre transform, Eq. (\ref{eq:GF_Legendre}):

\begin{equation}
G_{chain}(f_{tr})-G_{globule}(f_{tr})=F_{chain}(f_{tr})-F_{globule}(f_{tr})-f_{tr}(D_{chain}-D_{globule})\label{eq:Maxwell1}\end{equation}
Since \begin{equation}
F_{chain}(f_{tr})-F_{globule}(f_{tr})=\int_{D_{globule}}^{D_{chain}}\frac{\partial F}{\partial D}\, dD=\int_{D_{globule}}^{D_{chain}}f\, dD\end{equation}
then

\begin{equation}
G_{chain}(f_{tr})-G_{globule}(f_{tr})=\int_{D_{globule}}^{D_{chain}}(f-f_{tr})\, dD=0\label{eq:Maxwell3}\end{equation}
and this means that the area between the horizontal solid line and
the part of the dashed line above it is equal to the area between
the horizontal solid line and the part of the dashed line below it.
Fulfillment of the Maxwell rule in this case is a direct consequence
of the correctness of the Legendre transform.

The threshold force $f_{tr}$ is found to be smaller than the reaction
force at the onset of the stretched phase formation in the $D-$ensemble
and larger than the reaction force corresponding to disappearance
of the globular phase in the $D-$ensemble. In other words, in the
$f-$ensemble, a passage from the growing force branch at small deformations
(when the globule shape is close to that of a prolate ellipsoid) to
the unfolded open chain branch takes place at smaller deformation
and/or force compared to the onset of the microphase segregation in
the $D-$ensemble. This means that in the position-clamp mode a larger
maximum extension of the ellipsoidal globule can be reached compared
to the force-clamp mode. In the range of large deformations a similar
effect is observed: position-clamp mode allows obtaining less extended
open chain states than it is possible for the globule deformed in
the force-clamp mode.

In   Fig.~\ref{fig:comp_f_vs_D} one can find only one situation when deformation
curves totally coincide in both ensembles, this corresponds to $N=100$,
$\chi=0.8$ in   Fig.~\ref{fig:comp_f_vs_D} b. In this case, the globule
is deformed as a whole, its density decreases with extension and the
force monotonously increases with deformation (or, vice versa, the
deformation increases as the value of applied force grows).

\section{\label{sec:theory}Model and analytical theory of globule unfolding
in the $f$-ensemble}

SF-SCF analysis of the globule unfolding in the position-clamp mode
made in \cite{Polotsky:2009} allowed not only to obtain thermodynamic
characteristics of deformed globule and to calculate force-extension
curves but also to analyze changes in the structure of the macromolecule
upon increasing deformation. Using these findings, a simple model
of extended globule was introduced and analytical mean field theory
of the globule unfolding in the $D$-ensemble based on this model
was developed in\cite{Polotsky:2010:MM} that goes beyond the limits
of the SCF modeling. In present Section we extend the analytical approach
suggested in \cite{Polotsky:2010:MM} to the globule unfolding in
the $f-$ensemble. Our approach consists in calculating and then comparing
the free energies of the possible conformational states of the deformed
globule (1)-(3) shown in  Fig.~\ref{fig:FEC_and_conf}. Preliminary results
of this part of work were presented in \cite{Polotsky:2010:MS}.

\subsection{Weakly deformed globule}

Similarly to \cite{Polotsky:2010:MM} the shape of the unperturbed
globule (at $f=0$) formed by a macromolecule consisting of $N$ monomer
units and immersed into a poor solvent can be approximated by a sphere
of uniform density. The sphere has a volume $V$, polymer density
within the globule $\varphi=N/V$ is determined by the value of the
Flory parameter $\chi$ (i.e. by solvent quality). The radius of the
unperturbed globule is

\begin{equation}
R_{0}=\left(\frac{3N}{4\pi\varphi}\right)^{1/3}.\end{equation}

An important model assumption is the restriction imposed on possible
deformation of the globule. It is assumed that upon extension, the
globule undergoes only the shape deformation whereas its volume $V=4\pi R_{0}^{3}/3$
is conserved: the sphere acquires a prolate shape which we approximate
by a prolate uniaxial ellipsoid (often called spheroid), the major
axis of the ellipsoid is equal to the given end-to-end distance $D$.
Then the Helmholtz free energy of the the globule deformed in the
$D-$ensemble is \cite{Polotsky:2010:MM} 

\begin{equation}
F_{globule}=\mu N+\gamma S=\mu N+\gamma\cdot4\pi R_{0}^{2}\, g(x).\label{eq:Fglobule}\end{equation}
where $g(x)$ is a universal function of the extension parameter $x\equiv D/(2R_{0})$
\begin{equation}
g(x)=\frac{S}{4\pi R_{0}^{2}}=\frac{1}{2x}+\frac{1}{2}\frac{x^{2}}{\sqrt{x^{3}-1}}\arcsin\sqrt{\frac{x^{3}-1}{x^{3}}}\label{eq:g}\end{equation}

In Eq. (\ref{eq:Fglobule}) the first term describes the preference for
the monomer units to be in the globule, $\mu<0$ is the monomer chemical
potential in the globular phase. $\mu$ amounts to the monomer free
energy change when it is transferred from the pure solvent (dilute
phase, which is taken as a reference state for calculation of the
chemical potential) to the globular phase.

The second term accounts for energetic penalties at the globule surface
which are proportional to the surface area $S$, $\gamma>0$ is the
interfacial tension coefficient. In order to pass from the $D$- to
the $f-$ensemble, the Legendre transform, Eq. (\ref{eq:GF_Legendre}),
should be used. This gives for the Gibbs free energy:

\begin{equation}
G_{globule}=F_{globule}-f\cdot D=\mu N+\gamma\cdot4\pi R_{0}^{2}\, g(x)-f\cdot D,\label{eq:Gglobule}\end{equation}

Equilibrium end-to-end distance (ellipsoid major axis) $D$ at given
applied force is obtained by minimizing $G_{globule}$, Eq. (\ref{eq:Gglobule})
with respect to $D$ : $\partial G_{globule}/\partial D=0$. This
leads to equation that contains $D=D(f)$ in implicit form: \begin{equation}
2\pi R_{0}\gamma\, g'(x)=f.\label{eq:eq_globule}\end{equation}

Here $g'(x)$ is the $g(x)$ derivative with respect to $x$: 

\begin{equation}
g^{\prime}(x)=\frac{1}{2}\left[-\frac{1}{x^{2}}+\frac{3x}{2(x^{3}-1)}+\frac{x(x^{3}-4)}{2(x^{3}-1)^{3/2}}\arcsin\sqrt{\frac{x^{3}-1}{x^{3}}}\right]\label{eq:g_prime}\end{equation}

Note that in the $D$-ensemble the expression for the reaction force
$f=f(D)$ was obtained in \cite{Polotsky:2010:MM} according to $f=\partial F_{globule}/\partial D$
from Eq. (\ref{eq:Fglobule}). This expression coincides with Eq. (\ref{eq:eq_globule}).

Function $g'(x)$ is a non-monotonic bounded function: it increases
from zero at $x=0$, passes through a maximum at $x=x^{*}\approx$
2.1942 corresponding to $g'(x^{*})\equiv g^{*}\approx0.2214$, and
then decays with extension asymptotically tending to zero. Hence,
in the framework of given model one can adequately describe globule
extensions up to $D/(2R_{0})\simeq2.2$ which corresponds to applied
forces that do not exceed a certain value \begin{equation}
f\leq f^{*}=2\pi R_{0}\gamma g^{*}\label{eq:condition_f}\end{equation}
Let us also give approximate expressions in the limit of small extensions
that will be used in the following analysis:

\begin{equation}
\left.\begin{aligned}g(x) & \simeq1+\frac{2}{5}(x-1)^{2}\,\\
g'(x) & \simeq\frac{4}{5}(x-1)\,\end{aligned}
\right.\,,\quad x-1\ll1\label{eq:gg_appr}\end{equation}

By substituting Eqs. (\ref{eq:eq_globule}) and (\ref{eq:gg_appr}) into \ref{eq:Gglobule}),
approximate expression for the Gibbs free energy of the globule as
function of applied force in the limit of small forces is obtained:

\begin{equation}
G_{globule}\simeq\mu N+4\pi\gamma\cdot R_{0}^{2}\,-2R_{0}f-\frac{5f^{2}}{8\pi\gamma}.\label{eq:Gglobule_appr}\end{equation}
Correspondingly, approximate expression for $D$:

\begin{equation}
D_{globule}\simeq2R_{0}+\frac{5f}{4\pi\gamma},\label{eq:Dglobule_appr}\end{equation}

The model described above depends on the (partial) parameters $\varphi$
(polymer density within the globule), $\mu$ (monomer chemical potential
within the globule), and $\gamma$ (interfacial tension coefficient)
characterizing the globular state. In our model it is assumed that
they are virtually $N-$independent and can be found as functions
of the solvent quality only. In \cite{Polotsky:2010:MM} it was demonstrated
that the dependence of $\varphi$ and $\mu$ on $\chi$ can be easily
found in the framework of the Flory lattice model of polymer solution: 

\begin{equation}
\chi=-\frac{\log(1-\varphi a^{3})}{(\varphi a^{3})^{2}}-\frac{1}{\varphi a^{3}}\,,\end{equation}
\begin{equation}
\mu=2+\frac{2-\varphi a^{3}}{\varphi a^{3}}\log(1-\varphi a^{3})\,.\end{equation}
For the interfacial tension coefficient $\gamma$ a closed form analytical
solution was obtained for moderately poor solvent (close to the coil-globule
transition point):

\begin{equation}
\gamma a^{2}=\frac{3}{16}(1-2\chi)^{2}\end{equation}
Alternatively, the values of $\varphi$, $\mu$, and $\gamma$ were
calculated in \cite{Polotsky:2010:MM} from SCF modeling of free globules;
they are presented in Table 1. The numerical results for$\varphi$
and $\mu$ agree wery good with the Flory theory in the whole range
of $\chi$, whereas for $\gamma$ an agreement with analytical result
is observed, as expexted, only at small $\chi$. In the following
numerical calculations we will use the values of$\varphi$, $\mu$,
and $\gamma$ given in Table 1. Hence, the proposed theory of globule
unfolding is a \emph{quantitative} one.

\begin{table}[t]
\begin{tabular}{|c|c|c|c|c|c|c|c|}
\hline 
 & $\chi=0.8$  & $\chi=1.0$ & $\chi=1.2$ & $\chi=1.4$ & $\chi=1.6$ & $\chi=1.8$ & $\chi=2.0$\tabularnewline
\hline
\hline 
$\mu$ & -0.10 & -0.23 & -0.37 & -0.54 & -0.71 & -0.89 & -1.08\tabularnewline
\hline 
$\varphi a^{3}$ & 0.54 & 0.70 & 0.80 & 0.87 & 0.92 & 0.94 & 0.96\tabularnewline
\hline 
$\gamma a^{2}$  & 0.088 & 0.18 & 0.27 & 0.38 & 0.48 & 0.58 & 0.67\tabularnewline
\hline
\end{tabular}

\caption{Values of monomer chemical potential ($\mu$) polymer density ($\varphi$)
and interfacial tension coefficient ($\gamma$) calculated for different
$\chi$ using SCF approach \cite{Polotsky:2010:MM}.}

\end{table}

Dependences of $G/N$ for deformed globule, Eq. (\ref{eq:Gglobule}), as
functions of $f$ at $0\leq f\leq f^{*}$ for various $N$ and $\chi$
are shown below in   Fig.~\ref{fig:G_vs_force_theory}. Extreme right points
correspond to the values of $f^{*}$increasing with an increase in
$\chi$.

\subsection{Stretched open chain}

When the applied force is large, the globule is completely unfolded
and strongly stretched, all the monomer units are exposed to solvent.
In \cite{Polotsky:2009,Polotsky:2010:MM}, the Helmholtz free energy
$F_{chain}$ in the $D-$ensemble was calculated on the basis of the
Gibbs free energy $G_{chain}$ in the $f-$ensemble subjected to the
Legendre transform, Eq. (\ref{eq:GF_Legendre}). Expression for the Gibbs
free energy of a freely jointed chain consisting of $N$ monomer units
and subjected to the force $f$ has the following form: 

\begin{equation}
G_{chain}=-N\cdot\log\left\{ 1+\frac{1}{2k}\left[\cosh\left(fa\right)-1\right]\right\} ,\label{eq:Gchain}\end{equation}
where $k$ is the rigidity parameter of the chain. Eq. (\ref{eq:Gchain})
was obtained in \cite{Polotsky:2010:MM} for lattice model, and $k$
is associated with the statistics of corresponding lattice walks. 

The average end-to-end distance $D$ at given force $f$ is obtained
by differentiation of the Gibbs free energy, Eq. (\ref{eq:Gchain}), with
respect to the force:

\begin{equation}
D=-\frac{\partial G_{chain}}{\partial f}=Na\cdot\frac{\sinh\left(fa\right)}{2k+\cosh\left(fa\right)-1}\label{eq:Dchain}\end{equation}

Eqs. (\ref{eq:Gchain}) and (\ref{eq:Dchain}) contain $k$ as the only one
parameter. In \cite{Polotsky:2009,Polotsky:2010:MM} the value of
$k=3/4$ was accepted. Below in   Fig.~\ref{fig:G_vs_force_theory} a universal
dependence of $D/(Na)$ on $f$ is shown.

Let us also write down the approximate expressions for $G_{chain}$
and $D$ obtained for small forces via Taylor expansion:\begin{equation}
G_{chain}\simeq-N\:\frac{f^{2}a^{2}}{4k}\label{eq:Gchain_appr}\end{equation}
\begin{equation}
D_{chain}\simeq Na^{2}\:\frac{f}{2k}\label{eq:Dchain_appr}\end{equation}
Approximate expressions are in good agreement with the exact ones
only in a narrow force range: $0\leq fa\lesssim0.5$, the value of
average extension turns out to be overestimated.

\subsection{Microphase-segregated {}``tadpole'' state}

Consider now the third of possible {}``candidates'' which is the
mixed microphase-segregated tadpole state where globular and extended
part coexist in equilibrium.. Let us assume that the ellipsoidal head
comprises $n$ monomer units and the tail, therefore, consists of
$N-n$ units. Both the head and the tail are subjected to the force
$f$. Then the Gibbs free energy of the microphase-segregated structure
can be written using Eqs. (\ref{eq:g}), (\ref{eq:Gglobule}), and (\ref{eq:Gchain}):
\begin{equation}
G_{tadpole}=G_{globule}(n,\, f)+G_{chain}(N-n,\, f)\label{eq:Gtadpole}\end{equation}
(in expressions for $G_{globule}$, Eq. (\ref{eq:Gglobule}), and $G_{chain}$,
Eq. (\ref{eq:Gchain}), instead of $N$ as argument, $n$ and $N-n$, respectively,
are substituted). The size of the head, $n$, assures phase equilibrium
in the tadpole conformation, hence, the free energy $G_{tadpole}$
should be minimized with respect to $n$.

\begin{figure}[t]
\includegraphics[width=8cm]{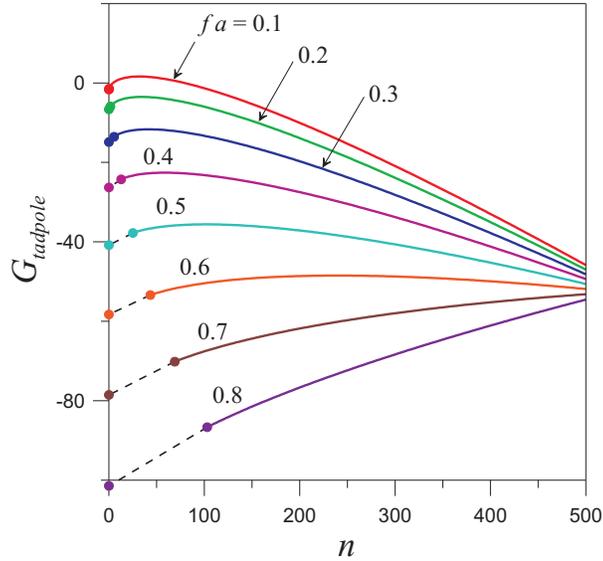}

\caption{\label{fig:G_tadpole}Gibbs free energy of the tadpole conformation
as function of the number of monomers in the globular head calculated
for $N=500$ , $\chi=1$ and various values of applied force $f$.}

\end{figure}

It was shown above, see   Fig.~\ref{fig:G_vs_force_SCF_200}, that according
to the SCF results the free energy of the two-phase state, $G_{tadpole}$,
is larger than the free energies of the one-phase states, $G_{globule}$
and $G_{chain}$ . Analytical theory provides us with opportunity
of deeper analysis of this issue.   Fig.~\ref{fig:G_tadpole} presents
as an example a series of dependences of the tadpole free energy on
the distribution of monomer units between two phases. It can be seen
that all $G_{tadpole}(n)$ curves are convex up, that is, $G_{tadpole}(n)$
has only boundary minima corresponding to pure states: globule at
$n=N$ or open chain at $n=0.$ Hence, in the equilibrium the microphase-segregated
tadpole state with $0<n<N$ is unstable and is not implemented. The
system will be in one of two pure states: weakly extended compact
globule at small forces or stretched open chain at large $f$. For
the particular choice of parameters in  Fig.~\ref{fig:G_tadpole} ($N=500$
, $\chi=1$), at $f\leq0.5$ the globule state with $n=N$ is stable
while at $f\geq0.6$ the free energy minimum corresponds to the open
chain state with $n=0$. Therefore, the transition between two states
occurs in the force range $0.5<f<0.6$. Note also that the curves
in   Fig.~\ref{fig:G_tadpole} have a gap at small $n$ between $n>0$
and $n=0$. This is because according to the condition (\ref{eq:condition_f})
small globular head becomes unstable if the applied force exceeds
the threshold value $f=f^{*}$ which depends on the number of monomer
units in the globular phase.

\subsection{Gibbs free energy as function of applied force}

In the preceding subsection it was shown that a microphase-segregated
tadpole state is not formed in the globule subjected to a pulling
force, the system can be found in one of two pure states: weakly extended
ellipsoidal globule or stretched open chain. Therefore, for the analysis
it suffices to consider the free energies of these two states. Note
that in the framework of the approach that we use, only
one state corresponding to the free energy minimum (or, equivalently,
the ground state) is physically implemented in the system at the given
value of the governing parameter (i.e. of the force).

\begin{figure}[t]
\includegraphics[width=8cm]{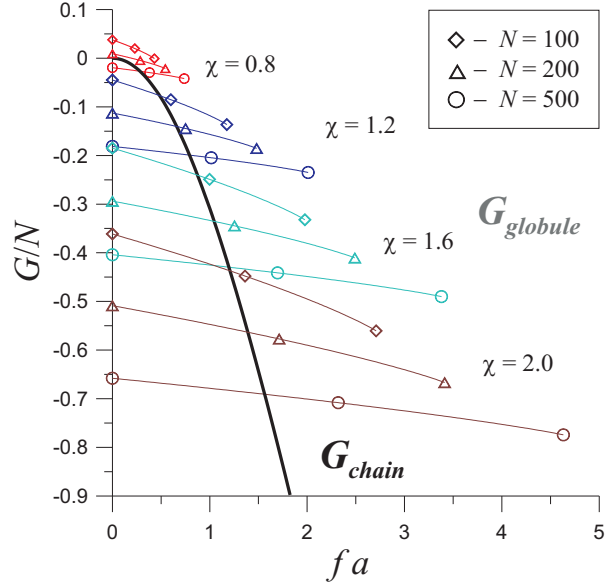}

\caption{\label{fig:G_vs_force_theory} Gibbs free energy per monomer unit
of compact globular and stretched states as function of applied force
for various values of $N$ and $\chi$ calculated using analytical
theory.}

\end{figure}

Fig.~\ref{fig:G_vs_force_theory} shows dependences of the reduced Gibbs
free energy $G/N$ of two pure state of stretched macromolecules on
applied force calculated using analytical theory for a series of $N$
and $\chi$ values. As follows from Eq. (\ref{eq:Gchain}), the monomer
free energy in the open chain state, $G_{chain}/N$, is a universal
function of the force independent of the chain length and the solvent
quality whereas the globule free energy per monomer, $G_{globule}/N$,
depends on both $N$ and the solvent quality. $G_{globule}/N$ dependences
are plotted in the range $0\leq f\leq f^{*}$, where $f^{*}$ is the
limiting maximum value of the force determined by Eq. (\ref{eq:condition_f}). 

One can see that at small applied forces $G_{globule}<G_{chain}$,
i.e. the global free energy minimum corresponds to the globule state.
With an increase in the value of $f$ the free energy difference for
two states, $G_{chain}-G_{globule}$, decreases, and at some value
$f=f_{tr}$ the curves intersect and $G_{globule}=G_{chain}$. At
$f>f_{tr}$, $G_{chain}<G_{globule}$ and the globule is unfolded.
Therefore, at $f=f_{tr}$ transition from the compact globule to the
unfolded state occurs, and this transition is a jumpwise first order
phase transition. With an increase of the Flory parameter $\chi$
(at fixed chain length) the position of the transition point moves
to the right: indeed, the more dense is the globule, the more effort
is needed to unfold it.

With a decrease in the chain length the value of the relative free
energy $G_{globule}/N$ increases which is easily explained: surface
area of the globule $\sim N^{2/3}$, correspondingly, the surface
contribution to the free energy per monomer unit is $\sim\gamma N^{-1/3}$,
i.e. it increases with decreasing $N$ (volume contribution to the
free energy per monomer unit is $N$-independent and equal to $\mu$).
Since $G_{chain}/N$ does not depend on $N$ this means that the larger
is $N$, the larger $f$ is required to unfold the globule. Increasing
$N$ also leads to broadening of the stability range for the globule
state $0\leq f\leq f^{*}$, in accordance with Eq. (\ref{eq:condition_f}).

It is interesting to note that for relatively small value of $\chi=0.8$
for $N=200$ and 100 formally calculated free energy of the globule
lies above the chain free energy in the whole $f$ range. This means that the macromolecule does not form a globule. 
This is a consequence
of the model simplifications. More accurate (and assumption-free) SCF calculations
show that at $\chi=0.8$ unfolding of small globules occurs according
to conventional scenario of the phase transition: there are two branches
that cross at the transition point. The difference is, however, that
extension of a small globule in a moderately poor solvent in the $D-$ensemble
occurs without formation of the microphase segregated tadpole state
but with a progressive decrease in the density of the globule \cite{Polotsky:2009}.

\subsection{Deformation curves (theory and SCF modeling)}

Let us compare deformation curves $D=D(f)$ obtained in the force-clamp
mode by SCF modeling and by using the analytical theory. In order
to be able to compare deformation curves corresponding to different
$N$ and $\chi$, not the absolute but the relative extension of the
macromolecule with respect to its contour length, $D/(Na)$, should
be considered. An example of such a comparison is presented in  Fig.~\ref{fig:D_vs_f_reduced}
which was already discussed above (deformation curves calculated in
SCF modeling are shown by solid lines, {}``theoretical'' curves
- by dashed lines). One can see that a good agreement is observed
for longer chain and for stronger solvent. At these values of parameters
the suggested model properly accounts for main peculiarities of the
globule behavior upon deformation in the force-clamp mode. While the
applied force is small enough, the globule is good described by an
ellipsoid of constant density, when the force reaches its threshold
value, the globule unfolds according to {}``all-or-none'' mechanism,
and the end-to-end distance abruptly increases. The value of the jump
grows with increasing $N$ and $\chi$. The threshold value of the
force obtained in the framework of the theory is somewhat lower than
that obtained from the SCF modeling. 

With an decrease in the chain length and/or solvent strength the simple
model fails to describe adequately the behavior of the globule. Note
that in  Fig.~\ref{fig:D_vs_f_reduced} a for $N=100$ $\chi=0.8$ and
1.0 and $N=200$, $\chi=0.8$ there are only curves obtained by SCF
modeling (solid lines) whereas corresponding theoretical curves are
missing. In all these cases globule deformation in the conjugated
$D-$ensemble occurs by strong extension and depletion (decrease in
the density) of the globule which cannot be taken into account in
our simple model. However, at $N=100$ and $\chi=1.0$, as well as
at $N=200$, $\chi=0.8$, $f(D)$ dependence in the $D$-ensemble
contains a part with {}``anomalous'' dependence, see  Fig.~\ref{fig:comp_f_vs_D},
hence, in the $f$-ensemble $G(f)$ dependence has a loop and globule
unfolding occurs with a jump (in this case {}``depleted'' states
are {}``hurdled''),   Fig.~\ref{fig:D_vs_f_reduced} a. At $N=100$ and
$\chi=0.8$ $f(D)$ dependence in the $D$-ensemble is monotonously
increasing which lead in the $f$-ensemble to monotonous $G(f)$ and
$D(f)$ dependences. Moreover $D(f)$ dependence in the $f$-ensemble
is obtained by simply inverting the $f(D)$ dependence in the $D$-ensemble.

The results here presented demonstrate a good agreement between SCF
modeling and analytical theory based on simple model. This correspondence
is seemed to us essential in two respects.

On the one hand, data of the SCF modeling were not obtained by directly
modeling (i.e. performing SCF calculations) globule unfolding in the
$f$-ensemble but were derived from the results of SCF modeling in
the $D$-ensemble using the principles of statistical physics. The
obtained agreement gives support to correctness of such an approach.

On the other hand, the observed correspondence demonstrates a wide
applicability range of the analytical approach and allows using the
results of the analytical theory for analysis of the system behavior.

\subsection{Transition point}

Analytical theory allows to carry out a more detailed study of the
unfolding transition point and of the changes in globule characteristics
at the transition. In the transition point, the free energy minima
corresponding to the globule and the open chain states have the same
depth and the position of the transition can be found, therefore,
from the following equation:

\begin{equation}
G_{globule}(N,\, f_{tr})=G_{chain}(N,\, f_{tr})\label{eq:TP}\end{equation}

\begin{figure}[t]
\includegraphics[width=8cm]{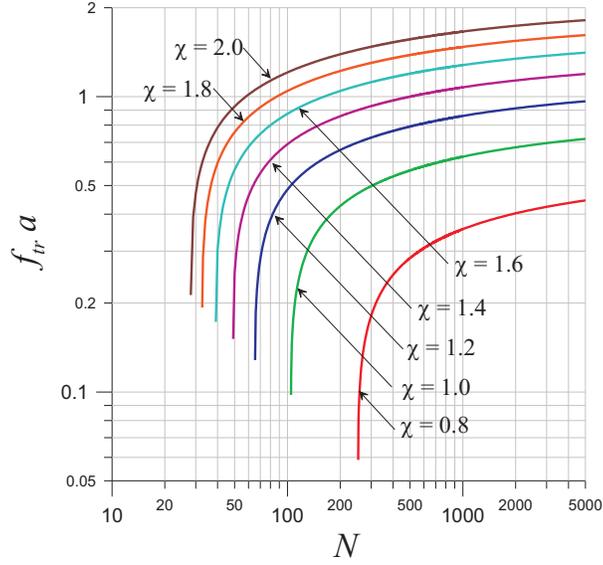}

\caption{\label{fig:ftr_vs_N}Threshold (transition) force in the $f$-ensemble
as function of polymerization degree at various values of $\chi$.}

\end{figure}

\begin{figure}[t]
\includegraphics[width=8cm]{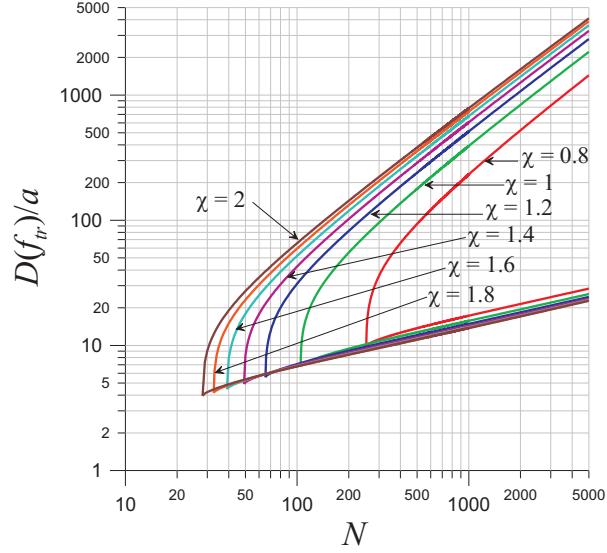}

\caption{\label{fig:Dtr_vs_N} Globule and open chain extension at threshold
(transition) force as function of polymerization degree at various
values of $\chi$. Lower branch corresponds to globule state, upper
branch - to open chain state.}

\end{figure}

Fig.~\ref{fig:ftr_vs_N} shows the dependence of the threshold force $f_{tr}$
on the polymerization degree $N$ at different values of the Flory
parameter $\chi$ while in   Fig.~\ref{fig:Dtr_vs_N} corresponding dependences
of the average end-to-end distance in globule (lower branches) and
open chain (upper branches) states calculated in the transition point
are presented. One can see that with an increase in $\chi$ and/or
$N$ both the threshold force, $f_{tr}$, and the jump in the end-to-end
distance, $D_{chain}(f_{tr})-D_{globule}(f_{tr})$, increase. At small
$N$, two branches corresponding to $D_{chain}$ and $D_{globule}$
meet at a certain point $N=N_{cr,f}(\chi)$ ($f_{tr}(N)$ also terminates
in this point). At $N<N_{cr,f}$ , Eq. (\ref{eq:TP}) has no solution (and
the model itself is not applicable at small $N$ where the crucial
assumption about conservation of the globule density $\varphi$ upon
extension does not work; this is also illustrated by  Fig.~\ref{fig:G_vs_force_theory}).
It can be concluded that $N=N_{cr,f}$ is a \emph{critical point for
the given model}.

To discuss the obtained dependences let us make use of approximate
expressions for the Gibbs free energy of the globule, Eq. (\ref{eq:Gglobule_appr}),
and the open chain, Eq. (\ref{eq:Gchain_appr}), at small applied forces
$fa\ll1$, and the corresponding expressions for the average end-to-end
distance, Eqs. (\ref{eq:Dglobule_appr}) and (\ref{eq:Dchain_appr}). By equating
the free energies of two states, a quadratic equation with respect
to $f$ is obtained:

\begin{equation}
\mu N+4\pi\gamma R_{0}^{2}-2R_{0}f+\frac{1}{4k}f^{2} a^2 N-\frac{5f^{2}a^{2}}{8\pi\gamma}=0\end{equation}

By keeping only leading in $N$ terms (of the order $N$ and $N^{2/3}$)
we obtain that

\begin{equation}
\begin{aligned}f_{tr} & \thickapprox\frac{2\sqrt{k\bigl|\mu\bigr|}}{a}\left(1-\Delta_{tr}\right),\end{aligned}
\label{eq:ftr_appr}\end{equation}

where

\begin{equation}
\Delta_{tr}=\left(\frac{9}{2}\right)^{1/3}\frac{\gamma}{\left|\mu\right|}\left(\frac{\pi}{\varphi^{2}N}\right)^{\frac{1}{3}}\end{equation}
is the first correction term of the order $1/N^{1/3}$. With an increase
in $N$ this term decreases and the threshold force increases tending
to the limiting (maximum) value $f_{tr}(N\to\infty)=2\sqrt{k\bigl|\mu\bigr|}/a$.
The value of the limiting force grows with $\chi$ due to $\mu(\chi)$
dependence. The correction term $\Delta_{tr}$weakly decreases with
an increase in $\chi$ (as follows from the data of Table 1). Using
Eqs. (\ref{eq:ftr_appr}), (\ref{eq:Dglobule_appr}), and (\ref{eq:Dchain_appr})
for the average end-to-end distance, asymptotic expressions for $D_{globule}$
and $D_{chain}$ in the transition point can be found:

\begin{equation}
D_{tr,\, globule}\simeq2R_{0}+\frac{5f_{tr}}{4\pi\gamma}\approx\left(\frac{6N}{\pi\varphi}\right)^{\frac{1}{3}}+\frac{5\sqrt{k\bigl|\mu\bigr|}}{2\pi\gamma}\left(1-\Delta_{tr}\right)\label{eq:Dtrgl_appr}\end{equation}

\begin{equation}
D_{tr,\, chain}\simeq Na^{2}\:\frac{f}{2k}\approx Na\,\sqrt{\frac{\left|\mu\right|}{k}}\left(1-\Delta_{tr}\right)\label{eq:Dtrch_appr}\end{equation}

As follows from Eq. (\ref{eq:Dtrgl_appr}), $D_{tr,\, globule}\sim N^{1/3}$
and weakly depends on $\chi$. The value of $D_{tr,\, chain}$ at
not too small $N$ and $\chi$ is proportional to $N$. The difference
between $D_{tr,\, globule}$ and $D_{tr,\, chain}$ demonstrates a
jump in the size in the transition point that grows with an increase
in $N$ and $\chi$.

\begin{figure}[t]
\includegraphics[width=8cm]{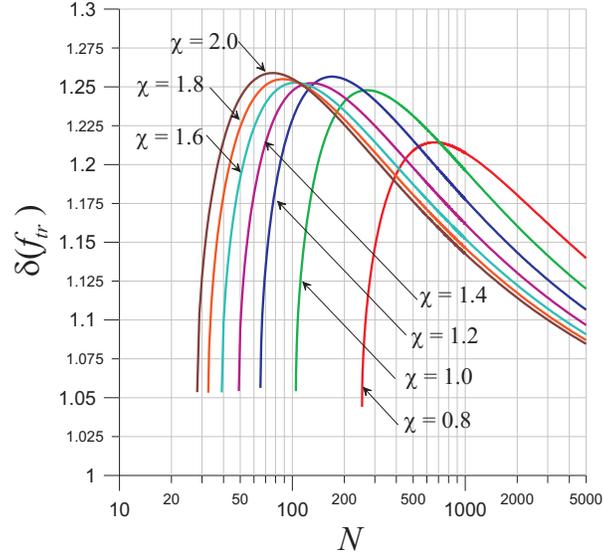}

\caption{\label{fig:astr_vs_N} Asymmetry of the globule in the transition
point as function of chain length at various values of $\chi$.}

\end{figure}

Consider now another characteristic of the globule in the transition
point - the asymmetry, or the long-to-short axis (longitudinal-to-transversal
size) ratio $\delta=D/(2R_{\bot})$ presented in   Fig.~\ref{fig:astr_vs_N}.
One can see that the asymmetry in the transition point is a non-monotonous
function of $N$: with an increase in $N$ it grows, passes through
a maximum and then decays. Position of the maximum depends on the
value of the Flory parameter: as $\chi$ increases, it shifts towards
smaller $N$. Note that the height of the maximum is a little larger
than 1.25 i.e. the shape of the globule in the transition point only
slightly deviates from the spherical one (when $\delta=1$).

The non-monotonicity of the function $\delta(N)$ follows also from
simple arguments. The asymmetry $\delta$ is related to the degree
of stretching of the globule provided that the globule volume is conserved.
As it was shown in \cite{Polotsky:2010:MM}, $\delta=x^{3/2}$, where
$x=D/(2R_{0})$ was used above in Eqs. (\ref{eq:g}), (\ref{eq:g_prime}),
and (\ref{eq:gg_appr}). It follows from Eq. (\ref{eq:Dtrgl_appr}) that

\begin{equation}
x=1+C_{1}N^{-1/3}(1-C_{2}N^{-1/3}),\end{equation}
where $C_{1}$and $C_{2}$ are $N$-independent functions of $\chi$.
It is easy to see that $x$ is a non-monotonous function of $N^{-1/3}$with
a maximum at $N^{1/3}=2C_{2}$.

\section{\label{sec:discussion}Discussion}

\subsection{Deformation curves and characteristics of the transition}

In  Fig.~\ref{fig:comp_f_vs_D} force-extension curves in two modes of
extension calculated by using SCF modeling are compared. Similar dependences
were calculated using the analytical theory (not presented here).
It was shown above how the complex deformation curve obtained in the
position-clamp mode is transformed into connected by a jump two branches
of the deformation curve in the force-clamp mode, see Fig.~\ref{fig:Legendre}
and Fig.~\ref{fig:comp_f_vs_D}. From the general thermodynamic consideration
it follows that the value of $f_{tr}$ at which the abrupt unfolding
transition takes place in the force-clamp mode (in the $f$-ensemble)
can be found from the Maxwell rule applied to the deformation curve
in the position-clamp mode (in the $D$-ensemble), see. Eqs. (\ref{eq:Maxwell1})-(\ref{eq:Maxwell3})
above. Therefore, if the globule is unfolded by applying a force,
the states corresponding to {}``anomalous'' part of the deformation
curve in the $D$-ensemble characterized by the force decay with increasing
extension $D$ are {}``hurdled''. It is this range of extensions
in the position-clamp mode where multistep globule unfolding occurs
including (i) formation of extended tail, (ii) growth of the latter
with extension in the phase coexistence regime and (iii) the abrupt
breakdown of the diminished globular head.

Let us compare the value of the threshold force $f_{tr}$ in the $f$-ensemble
with characteristic values of the reaction force in the $D$-ensemble:
the force $f_{1}$ at the extension $D_{1}$ corresponding to the
onset of extended phase formation (i.e. to formation of the tadpole)
and the force $f_{2}$ at the extension $D_{2}$ where the globular
phase disappears. Analytical expressions for $f_{1}$ and $f_{2}$
were derived in \cite{Polotsky:2010:MM} in the same approximation
as Eqs. (\ref{eq:ftr_appr}) - (\ref{eq:Dtrch_appr}) and have the following
form:

\begin{equation}
f_{i}=\frac{2\sqrt{k\bigl|\mu\bigr|}}{a}\left(1-\Delta_{i}\right),\label{eq:f12_appr}\end{equation}
where $i=1,\,2$ and

\begin{equation}
\Delta_{i}=K_{i} \left[\frac{\gamma}{\bigl|\mu\bigr|}\left(\frac{\pi}{\varphi^{2}N}\right)^{\frac{1}{3}}\right]^{\alpha_{i}}<1.\label{eq:Delta_i}\end{equation}
Here $\alpha_{1}=1$, $\alpha_{2}=\frac{3}{4}$, and $K_{1}=(4/3)^{1/3}$
, $K_{2}=3^{3/4}2^{1/4}$. Comparing Eq. (\ref{eq:f12_appr}) and Eq. (\ref{eq:ftr_appr})
one can see that $f_{tr}$, $f_{1}$, and $f_{2}$ differ only in
$\Delta_{i}$. It is easy to check that $\Delta_{1}<\Delta_{tr}<\Delta_{2}$
and, correspondingly, $f_{1}>f_{tr}>f_{2}$ . Indeed, $\Delta_{1}$
and $\Delta_{tr}$ differ only in numerical coefficients, hence, the
left part of inequality is obvious. For $\Delta_{tr}$ and $\Delta_{2}$
one has

\begin{equation}
\frac{\Delta_{tr}}{\Delta_{2}}=\frac{1}{3^{1/3}2^{2/3}}\Delta_{2}^{1/3}<1\end{equation}
The latter inequality follows from the observation that $\Delta_{2}<1$.

The end-to-end distances $D_{1}$ and $D_{2}$ in pure globular and
extended phases are related to reaction forces $f_{1}$ and $f_{2}$
by expressions similar to Eqs. (\ref{eq:Dtrgl_appr}) and (\ref{eq:Dtrch_appr})
via changing $f_{tr}$ and $\Delta_{tr}$to $f_{1}$, $\Delta_{1}$
and $f_{2}$, $\Delta_{2}$, respectively. From the inequality connecting
$f_{tr}$, $f_{1}$, and $f_{2}$ , it follows that 

\begin{equation}
\begin{aligned}D_{tr}(globule)<D_{1}(globule)\\
D_{tr}(chain)>D_{2}(chain)\end{aligned}
.\end{equation}

\subsection{Phase diagrams}

\begin{figure}[t]
\includegraphics[width=8cm]{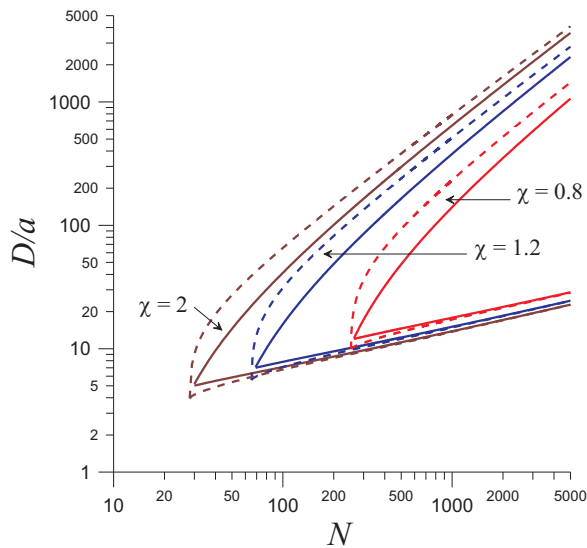}

\caption{\label{fig:comp_D_vs_N}Globule and open chain extension at threshold
force, $D_{tr}(globule)$ and $D_{tr}(chain)$, in the $f$-ensemble
(solid lines) and phase coexistence boundaries, $D_{1}(globule)$
and $D_{2}(chain)$, in the $D$-ensemble (dashed lines) as functions
of polymerization degree calculated at different $\chi$values. Lower
branches correspond to globule, upper branches - to unfolded open
chain state.}

\end{figure}

In   Fig.~\ref{fig:comp_D_vs_N} phase diagrams of deformed globule in
two modes of deformation are presented in $(N,D)$ coordinates for
various values of $\chi$. For the $f$-ensemble, upper and lower
branches of the diagram are determined by the values of $D_{tr}(globule)$
and $D_{tr}(chain)$ calculated in the transition point. For the $D$-ensemble
the branches are the phase coexistence boundaries $D_{1}(globule)$
and $D_{2}(chain)$. Both diagrams have similar shape (both upper
and lower $D$'s grow with an increase in $N$), phase diagrams for
the $D$-ensemble lie completely within corresponding diagrams for
the $f$-ensemble. In both ensembles the area below the lower boundary
corresponds to globular state whereas the area above the upper boundary
- to completely unfolded {}``open'' state. However, an essential
difference between the diagrams is that the area inside the diagram
in the $D$-ensemble corresponds to real stable microphase segregated
tadpole state while in the $f$-ensemble these state are unattainable.
The {}``difference'' between diagrams (i.e. area between solid and
dashed lines) are related to the above discussed pure states of {}``strongly
stretched globule'' (below) and {}``weakly stretched open chain''
(above) which can be obtained only in the position-clamp mode. 

At small $N$, diagrams for both deformation modes have a critical
point, but the position of the latter is slightly different in two
ensembles. For the globule in the $D$-ensemble the critical point
corresponds to the smallest chain length (for a given solvent quality)
which makes microphase segregation within the globule (i.e. formation
of the tadpole structure) possible. In the critical point the range
of extensions where the tadpole structure is stable {[}$D_{1}(globule);$$D_{2}(chain)${]}
degenerates into a point. On the left of the critical point in the
$D$-ensemble deformation of the globule gives rise to its gradual
elongation accompanied by the decrease in density.

The critical point in the force-clamp mode where the globule phase
diagram ($f_{tr}(N)$ curve,   Fig.~\ref{fig:ftr_vs_N}) ends and the jump
in the end-to-end distance $D_{tr}(chain)-D_{tr}(globule)$ disappears
is a critical point \emph{for our particular model} of the globule
unfolded by applied force. We recall that phase diagrams for two modes
of extension presented in   Fig.~\ref{fig:comp_D_vs_N} were calculated
in the framework of the simple model assuming extension-independence
of the globule density. A true critical point in the $f$-ensemble
must correspond to shorter chains, as it follows from the results
of SCF modeling, because $f(D)$ dependence in the $D$-ensemble remains
non-monotonous even if intramolecular microphase segregation does
not occur and globule is deformed {}``as a whole'', with decreasing
density, if we are not far from the critical point in the $D$-ensemble.
Such a non-monotonicity becomes apparent through {}``anomalous''
$f=f(D)$ dependence in the position-clamp mode corresponding to the
decrease of $D$ with increasing $f$. In the force-clamp mode this
range of extensions is always {}``hurdled''. In the true critical
point for the $f$-ensemble $N_{cr,f}(\chi)$ this non-monotonicity
disappear and at $N<N_{cr,f}$ deformation curves for both ensembles
coincide.

\section{\label{sec:conclusions}Conclusions}

We have presented a theoretical study of homopolymer globule unfolding
by a force applied to the ends of the macromolecule forming the globule,
i.e. in the force-clamp mode of extension equivalent to thermodynamic
$f$-ensemble. Two approaches were used to solve the problem: first,
the results of SF-SCF modeling of globule unfolding in the conjugate
$D$-ensemble, or, equivalent, in the position-clamp mode, were {}``translated''
into the $f$-ensemble. Namely, the dependence of the Helmholtz free
energy and the reaction force on the imposed end-to-end distance in
the $D$-ensemble were transformed according to the Legendre transform,
Eq. (\ref{eq:GF_Legendre}), to obtain the Gibbs free energy and equilibrium
end-to-end distance as functions of applied force in the $f$-ensemble,
the scheme of this {}``translation'' is shown in  Fig.~\ref{fig:Legendre}.
On the other hand, analytical mean-field theory of the globule unfolding
in the force-clamp mode was developed by using simple model of deformed
globule introduced in \cite{Polotsky:2010:MM}. The theory makes it
possible to go beyond the limits of the SCF calculations: first of
all, it allows calculations for large $N$, where the system size
is large and numerical SCF calculations become very time and memory
consuming. Moreover, in the framework of the developed theory it is
easy to calculate the transition point and to find corresponding conformational
characteristics in a wide range of $N$ and $\chi$.

One of the main goals of the study was to compare the globule unfolding
in two modes of extension. Our analysis, like all previous analysis
of this problem, shows that the globule behaves essentially differently
in these two situations. In the position-clamp mode, the globule unfolding
upon an increase in the end-to-end distance occurs via formation of
the microphase-segregated tadpole conformation with globular head
and stretched tail coexisting in equilibrium in a wide range of extension.
As the extension increases, the tail size grows, the head diminishes
and this is accompanied by {}``anomalous'' decrease of the reaction
force with extension. However, when the head contains approximately
$\sim N^{3/4}$ monomer units, it loses its stability and disintegrates
jumpwise As a result the force-extension curve has a complex structure.
This behavior in the position-clamp mode is caused by the small size
of the system under consideration. Polymer globule is a typical example
of a nanosystem where surface effects play a significant role. Additional
circumstance is the linear memory - connectivity of monomer units
into a chain.

In the force-clamp mode no such peculiarities are manifested. Mechanical
unfolding of a globule by applied force occurs without intramolecular
microphase segregation (formation of microphase-segregated tadpole
state): at certain threshold value of the pulling force the globule
unfolds as a whole. This transition (of ‘‘all or none’’ type) is accompanied
by a jump in the end-to-end distance. The values of both the threshold
force and of the jump in the end-to-end distance in the transition
point grow with an increase in $N$ and/or solvent strength.

In our previous work \cite{Polotsky:2010:MM}, statistical-mechanical
analogy between globule unfolding in the $D$-ensemble and liquid-gas
transition in the van der Waals gas in the $(V,T)$-ensemble below
the critical temperature, $T<T_{cr}$, was discussed. It was shown
that in spite of the obvious similarity between phase states (globular
phase $\leftrightarrow$ liquid, stretched (unfolded) phase $\leftrightarrow$
gas), governing parameters (end-to-end distance $D$ $\leftrightarrow$
volume $V$) and rearrangements in the systems upon increase in the
governing parameter {[}dense phase (globular or liquid) $\to$ phase
coexistence $\to$low density phase (unfolded or gas){]}, comparison
of $f=f(D)$ and $p=p(V)$ curves indicate a marked difference in
the behavior of small (globule) and macroscopic (gas) systems which
is observed in the phase coexistence regime. In the macrosystem {[}$(V,T)$-ensemble{]}
the pressure $p$ is constant at any volume where two phases coexist
while in the extended small globule the reaction force dependence
on $D$ is “anomalous”: an increase in $D$ in the phase coexistence
regime is accompanied by a decrease in $f$. In the asymptotic (thermodynamic)
limit $N\to\infty$, however, the reaction force in the phase coexistence
regime is constant (i.e. in Eqs. (\ref{eq:f12_appr})-(\ref{eq:Delta_i}) $\Delta_{1}=\Delta_{2}=0$
and $f_{1}=f_{2}$). 

In the $f$-ensemble (force-clamp mode) considered in the present
paper, the small system (globule) changes its phase state jumpwise
with an increase in applied force, similarly to what occurs in macroscopic
system in the $(p,T)$-ensemble. The state curves for the macroscopic
system in $(p,V)$ coordinates coincide in conjugate ensembles while for the globule, the
state (i.e. deformation) curves differ in conjugate ensembles and satisfy Maxwell area rule
[Fig.~\ref{fig:comp_f_vs_D} and Eqs. (\ref{eq:Maxwell1})-(\ref{eq:Maxwell3})],
the latter gives the value of the threshold force $f_{tr}$ corresponding
to the all-or-none transition in the $f$-ensemble.

Comparison of the deformation curves in two ensembles also shows that
not only the microphase segregated tadpole state but also some of
the pure states that are stable in the $D$-ensemble cannot be accessed
in the $f$-ensemble. This concerns ‘‘strongly extended globules’’
(close to the ellipsoid-tadpole transition point in the $D$-ensemble)
and ‘‘weakly stretched open chains’’ (just after the tadpole-open
chain {[}i.e. complete unfolding{]} transition point in the $D$-ensemble).
Or, in other words, the stretched phase is obtained at larger extensions
in the position-clamp mode than in the force-clamp mode. Moreover,
in the former case only a portion of the macromolecule is in the stretched
phase whereas in the latter case the whole macromolecule is unfolded
and stretched. Similarly, formation of a globule upon a decrease in
the extension of a strongly stretched chain (i.e. folding of the macromolecule
into a globule) in the $f$-ensemble occurs easier (at larger extensions)
than a nucleation of the globular phase in the $D$-ensemble.

We have shown that the jump in the globule unfolding in the force-clamp
mode is directly related to the {}``anomalous'' part on the force-extension
dependence in the position-clamp mode: the states corresponding to
the force decay are unstable (unfavorable) in the $f$-ensemble and
the system simply jumps over these states. When $f=f(D)$ curve in
the $D$-ensemble is monotonously increasing, globule unfolds similarly
in two modes, deformation curves coincide, and this case corresponds
to the pre-critical regime $N<N_{cr,f}(\chi)$, where $N_{cr,f}$
is the critical point for the $f$-ensemble, or the minimal chain
length necessary for the jump-wise unfolding of the globule. The value
of $N_{cr,f}$ is slightly less than the critical point for the $D$-ensemble,
$N_{cr,D}$, a minimal chain length below which the intramolecular
microphase segregation in the extended globule does not occur and
the globule is deformed “as a whole”, without intramolecular segregation
but by progressive “dissolution” of its core.

Let us mention that the developed theory may also help in understanding
the unfolding of more complex globular structures. Indeed, in the
position-clamp mode, force-extension curves of both homopolymer globule
and globular protein are non-monotonous consisting of ascending and
“anomalous” descending force branches. In the homopolymer case there
is one “anomalous” part whereas on the protein deformation curves
– they are multiple, separated by ascending parts, thus leading to
appearance of the “sawtooth” pattern (Fig.~\ref{fig:schem_dc_protein}
a). Drawing an analogy with the homopolymer globule unfolding, we
can relate each decreasing part of the protein globule force-extension
curve with unraveling of individual domains of the protein. In the
force-clamp mode, deformation (extension vs force) curves of homopolymer
and protein globules have a “staircase” structure. The homopolymer
globule “staircase” has only one step corresponding to jumpwise complete
unfolding of the globule. Deformation curve of globular protein has
several steps, each one is related to unfolding of individual protein
domains. Therefore, one can speak about universal features of both
simple (homopolymer) and complex globule unfolding in the $D$- and
$f$-ensembles.

Finally, let us recall that both the theory we developed and the SCF approach are of the mean-field type that neglect fluctuations
around the ground state for the considered system. On the other hand,
proper account of fluctuations and exact calculation of the partition
function of a deformed macromolecule may help in revealing fine differeces
in behavior of finite-length polymer chain deformed in $D$- and $f$-ensembles.
There is an ongoing discussion in the literature about possible inequivalence
of statistical ensembles in stretching of individual macromolecules,
even in simplest ``minimal'' polymer models such as Gaussian \cite{Neumann:2003,Suezen:2009}
and semiflexible \cite{Sinha:2005} chains in athermal solvent. A
comprehensive critical review of these and related works can be found
in \cite{Winkler:2010}. In the case of the globule deformation considered
in the present paper, marked differences in deformation behavior in
two ensembles are clearly seen already on the mean-field level. Taking
into account fluctuations should lead to “smoothening” of the transition
– instead of a jump, the transition should occur in a narrow interval
of the governing parameter. This issue will be considered in detail
in a forthcoming work, in particular, attention will be devoted to
fluctuations in the distribution of monomer units between globular
and stretched phases in deformed globule.

\section*{Acknowlegdement}
Financial support by the Russian Foundation for Basic Research (RFBR) through Project 11-03-00969-a and by the Department of Chemistry and Marterial Science of the Russian Academy of Sciences is gratefully acknowledged. 

\bibliographystyle{unsrt}
\bibliography{Globule}

\begin{thebibliography}{10}

\bibitem{Forman:2007}
J~R Forman and J~Clarke.
\newblock Mechanical unfolding of proteins: insights into biology, structure
  and folding.
\newblock {\em Curr. Opin. Struct. Biol.}, 17(1):58 -- 66, 2007.

\bibitem{Borgia:2008}
A~Borgia, P~M Williams, and J~Clarke.
\newblock Single-molecule studies of protein folding.
\newblock {\em Annu. Rev. Biochem.}, 77(1):101--125, 2008.

\bibitem{Puchner:2009}
E~M Puchner and H~E Gaub.
\newblock Force and function: probing proteins with afm-based force
  spectroscopy.
\newblock {\em Curr. Opin. Struct. Biol.}, 19(5):605 -- 614, 2009.

\bibitem{Williams:2002}
M~C Williams and I~Rouzina.
\newblock Force spectroscopy of single dna and rna molecules.
\newblock {\em Curr. Opin. Struct. Biol.}, 12(3):330 -- 336, 2002.

\bibitem{Moffitt:2008}
J~R Moffitt, Y~R Chemla, S~B Smith, and C~Bustamante.
\newblock Recent advances in optical tweezers.
\newblock {\em Annu. Rev. Biochem.}, 77(1):205--228, 2008.

\bibitem{Meglio:2009}
A~Meglio, E~Praly, F~Ding, J.-F Allemand, D~Bensimon, and V~Croquette.
\newblock Single dna/protein studies with magnetic traps.
\newblock {\em Curr. Opin. Struct. Biol.}, 19(5):615 -- 622, 2009.

\bibitem{Samori:2005}
B~Samori, G~Zuccheri, and P~Baschieri.
\newblock Protein unfolding and refolding under force: Methodologies for
  nanomechanics.
\newblock {\em ChemPhysChem}, 6(1):29--34, 2005.

\bibitem{Oberhauser:2001}
A~F Oberhauser, P~K Hansma, M~Carrion-Vazquez, and J~M Fernandez.
\newblock {Stepwise unfolding of titin under force-clamp atomic force
  microscopy}.
\newblock {\em Proc. Natl. Acad. Sci. USA}, 98(2):468--472, 2001.

\bibitem{Schlierf:2007}
M~Schlierf, H~Li, and J~M Fernandez.
\newblock {The unfolding kinetics of ubiquitin captured with single-molecule
  force-clamp techniques}.
\newblock {\em Proc. Natl. Acad. Sci. USA}, 101(19):7299--7304, 2004.

\bibitem{Balescu:1975}
R.~Balescu.
\newblock {\em {Equilibrium and Nonequilibrium Statistical Mechanics}}.
\newblock Wiley-Blackwell, 1975.

\bibitem{Skvortsov:2009}
A~M Skvortsov, L~I Klushin, and T~M Birshtein.
\newblock {Stretching and Compression of a Macromolecule under Different Modes
  of Mechanical Manipulations}.
\newblock {\em Polym. Sci. Ser. A. (Russia)}, 51:723--746, 2009.

\bibitem{Schroedinger:1944}
E~{Schr\"odinger}.
\newblock {\em {What is Life?}}
\newblock Cambridge University Press, Cambridge, 1944.

\bibitem{Halperin:1991:EL}
A~Halperin and E~B Zhulina.
\newblock {On the Deformation Behavior of Collapsed Polymers}.
\newblock {\em Europhys. Lett.}, 15:417, 1991.

\bibitem{Halperin:1991:MM}
A~Halperin and E~B Zhulina.
\newblock {Stretching Polymer Brush in Poor Solvent}.
\newblock {\em Macromolecules}, 24:5393--5397, 1991.

\bibitem{Cooke:2003}
R~Cooke and D~R~M Williams.
\newblock {Stretching Polymers in Poor and Bad Solvents: Pullout Peaks and an
  Unraveling Transition}.
\newblock {\em Europhys. Lett.}, 64:267--273, 2003.

\bibitem{Craig:2005:1}
A~Craig and E~M Terentjev.
\newblock {Stretching globular polymers. I. Single chains}.
\newblock {\em J. Chem. Phys.}, 122:194901, 2005.

\bibitem{Polotsky:2009}
A~A Polotsky, M~I Charlaganov, F~A~M Leermakers, M~Daoud, O~Borisov, and T~M
  Birshtein.
\newblock {Mechanical unfolding of a homopolymer globule studied by
  self-consistent field modelling}.
\newblock {\em Macromolecules}, 42:5360--5371, 2009.

\bibitem{Polotsky:2010:MM}
A~A Polotsky, M~Daoud, O~Borisov, and T~M Birshtein.
\newblock {A Quantitative Theory of Mechanical Unfolding of a Homopolymer
  Globule}.
\newblock {\em Macromolecules}, 43:1629--1643, 2010.

\bibitem{Rayleigh:1882}
L~Rayleigh.
\newblock {On the Equilibrium of Liquid Conducting Masses charged with
  Electricity}.
\newblock {\em Phylos. Mag.}, 14:184--186, 1882.

\bibitem{Lifshitz:1968}
I~M Lifshitz.
\newblock {Some problems of the statistical theory of biopolymers}.
\newblock {\em Sov.Phys.JETP}, 55:2408--2422, 1968.

\bibitem{GrosbergKhokhlov:1994}
A~Yu Grosberg and A~R Khokhlov.
\newblock {\em {Statistical Physics of Macromoecules}}.
\newblock AIS Press, New York, 1994.

\bibitem{Landau:5}
L~D Landau and E~M Lifshitz.
\newblock {\em {Statistical Physics}}.
\newblock Nauka, Moscow, 1976.

\bibitem{Sinha:2005}
S~Sinha and J~Samuel.
\newblock {Inequivalence of statistical ensembles in single molecule
  measurements}.
\newblock {\em Phys. Rev. E}, 71(2):021104, Feb 2005.

\bibitem{Winkler:2010}
Roland~G. Winkler.
\newblock Equivalence of statistical ensembles in stretching single flexible
  polymers.
\newblock {\em Soft Matter}, 6:6183--6191, 2010.

\bibitem{Polotsky:2010:MS}
A~A Polotsky, E~E Smolyakova, O~V Borisov, and T~M Birshtein.
\newblock {Mechanical Unfolding of a Homopolymer Globule: Applied Force versus
  Applied Deformation}.
\newblock {\em Macromol. Symp.}, 296:639--646, 2010.

\bibitem{Neumann:2003}
R~M Neumann.
\newblock {On the Precise Meaning of Extension in the Interpretation of
  Polymer-Chain Stretching Experiments}.
\newblock {\em Biophysical Journal}, 85(5):3418 -- 3420, 2003.

\bibitem{Suezen:2009}
M~S\"uzen, M~Sega, and C~Holm.
\newblock {Ensemble inequivalence in single-molecule experiments}.
\newblock {\em Phys. Rev. E}, 79(5):051118, May 2009.

\end{thebibliography}

\end{document}